\documentclass[manuscript]{aastex}

\usepackage{epsfig}             
\usepackage{graphicx}           
\usepackage{amssymb}            
\usepackage{color}              
\usepackage{url}                

\newcommand{\angeo}{\textit{Ann.~Geophys.}}

\newcommand{\jastp}{\textit{J.~Atmos.~Solar Terr.~Phys.}}

\newcommand{\Rsun}{\,\textrm{R}_{\odot}}
\newcommand{\thatis}{{i.e.}}
\newcommand{\kmps}{\,\mbox{km\,s$^{-1}$}}
\newcommand{\mpss}{\,\mbox{m\,s$^{-2}$}}
\newcommand{\Ang}{\,\textrm{\AA}}

\begin{document}

\title{Determination of Acceleration from 3D Reconstruction of Coronal Mass Ejections Observed by STEREO}

\author{Anand D. Joshi\altaffilmark{1} and Nandita Srivastava\altaffilmark{1}}
\affil{Udaipur Solar Observatory, Physical Research Laboratory,
P.O.~Box 198, Badi Road, Udaipur 313001, India}
\email{janandd@prl.res.in}

\begin{abstract}
We employ a three-dimensional (3D) reconstruction technique, for the first time to study the kinematics of six coronal mass ejections (CMEs), using images obtained from the COR1 and COR2 coronagraphs on board the twin STEREO spacecraft, as also the eruptive prominences (EPs) associated with three of them using images from the Extreme UltraViolet Imager (EUVI). A feature in the EPs and leading edges (LEs) of all the CMEs was identified and tracked in images from the two spacecraft, and a stereoscopic reconstruction technique was used to determine the 3D coordinates of these features. True velocity and acceleration were determined from the temporal evolution of the true height of the CME features. Our study of kinematics of the CMEs in 3D reveals that the CME leading edge undergoes maximum acceleration typically below $2\Rsun$. The acceleration profiles of CMEs associated with flares and prominences exhibit different behaviour. While the CMEs not associated with prominences show a bimodal acceleration profile, those associated with prominences do not. Two of the three associated prominences in the study show a high and rising value of acceleration up to a distance of almost $4\Rsun$ but acceleration of the corresponding CME LE does not show the same behaviour, suggesting that the two may not be always driven by the same mechanism. One of the CMEs, although associated with a C-class flare showed unusually high acceleration of over $1500\mpss$. Our results therefore suggest that only the flare-associated CMEs undergo residual acceleration, which indicates that the flux injection theoretical model holds good for the flare-associated CMEs, but a different mechanism should be considered for EP-associated CMEs.

\end{abstract}



\section{Introduction}\label{S:intro} 

Coronal mass ejections (CMEs) result from a loss of equilibrium in the magnetic configuration in the solar corona \citep{Priest1988,Klimchuk2001}. Several factors like, flux emergence, flux cancellation, reconnection, shear, etc., are thought to be responsible for this loss of equilibrium \citep{Forbes.etal2006,Seaton.etal2011}. Once the equilibrium is lost, the energy needed by the CME for its propagation is derived from the surrounding magnetic field \citep{Forbes2000,Low2001}. Very often, the energy of the surrounding field is sufficient not only to propel a CME, but also to accelerate it \citep{Alexander2006}.

\citet{Zhang.Dere2006} have categorised the evolution of CMEs into a three-phase process involving initiation, acceleration and propagation (Figure 1 in their paper). According to \citet{Zhang.Dere2006}, the initiation phase is the phase of slow rise of CMEs, and in the acceleration phase they undergo a very rapid increase in their velocity, while, in the propagation phase, the CME velocity remains more or less constant, \thatis, it experiences almost zero acceleration. Using LASCO \citep{Brueckner.etal1995} observations on board the SoHO spacecraft \citep{Domingo.etal1995a}, \citet{Yashiro.etal2004} have observed that the CME velocity in the outer corona varies from less than $100\kmps$ to over $3000\kmps$. The propagation of CMEs can be understood if we consider the forces acting on them, which are the Lorentz force, gravitational force, and drag because of the ambient solar wind. Of the three forces, the drag force is the strongest beyond a few solar radii, and the other two can be neglected \citep{Gopalswamy.etal2001b,Cargill2004,Vrsnak.etal2010}. This is further supported by results obtained by \citet{Gopalswamy.etal2000}. They have observed that although initial CME speeds range from $124-1056\kmps$, the speeds of the corresponding interplanetary ejecta are found to lie in the range of $320-650\kmps$, which is more or less the speed of the ambient solar wind. \citet{Cargill2004} have reported that speeds of interplanetary CMEs (ICMEs) corresponding to CMEs with speeds ranging from $100-2000\kmps$, as measured from coronagraphs, lie within $100-200\kmps$ of the ambient solar wind. However, the time a CME takes to reach the Earth, the transit time, is known to vary from less than a day to over four days. This indicates that most of the CME dynamics occurs closer to the Sun. \citet{Vrsnak.etal2010} have reported that transit times of broad, low-mass CMEs depend mainly on the surrounding solar wind speed, while those of narrow, massive CMEs depend mainly on the initial speeds of the CMEs. Recently, \citet{Manoharan.Rahman2011} have also found that most of the ICMEs tend to attain speeds close to that of the ambient solar wind, and have estimated travel times of the CMEs to reach a distance of \mbox{1\,AU} based on the CME initial speed and drag due to solar wind.

CMEs have been classified on the basis of their source regions. \citet{Gosling.etal1976}, using the coronagraph on Skylab spacecraft, were the first ones to report that CMEs associated with flares are faster than those associated with prominences. This was supported by observation of CMEs by \citet{MacQueen.Fisher1983} who used the \textit{K}-coronameter at Mauna Loa Solar Observatory. In addition, they also observed that the former type showed smaller acceleration with increase in height than the latter. \citet{Sheeley.etal1999} have also reported a similar result based on their technique to track features observed in SoHO/LASCO coronagraphs. \citet{Moon.etal2002} in a statistical study involving over 3200 CMEs observed from SoHO/LASCO have reported that flare-associated CMEs have a higher median speed than those associated with EPs. Their study also found that although the median acceleration of all the events is zero, it decreases a little for CMEs with high speeds ($>500\kmps$). \citeauthor{Srivastava.etal1999a} (\citeyear{Srivastava.etal1999a}; \citeyear{Srivastava.etal2000}) have found that gradual CMEs attain the speed of the ambient solar wind at about $20\Rsun$ from the Sun. Results from \citet{Gopalswamy.etal2001b} also are consistent with this study, who reported deceleration as high as $-100\mpss$ for fast CMEs (speed $>900\kmps$) from a combined study of SoHO/LASCO and radio observations from Wind spacecraft.

\sloppy \citet{Chen.Krall2003} have studied acceleration of three CMEs using SoHO/LASCO observations, and proposed that CME acceleration occurs in two phases, the `main' phase and the `residual' phase. While most of the acceleration occurs in the main phase, there lies a second phase of acceleration known as the residual acceleration in the outer corona. \textbf{\citet{Chen.Krall2003} and \citet{Chen.etal2006} have identified the main acceleration phase as the interval over which Lorentz force is the most dominant, while during the residual phase, Lorentz force is comparable to the two other force, viz, gravity and drag.} They have employed the magnetic flux rope model \citep{Chen1989} to show a relation between the height at the peak of main acceleration phase, and the  footpoint separation of the CME flux rope. In their model, \citet{Chen.Krall2003} have proposed that a change in duration of the flux injection \citep{Krall.etal2000} determines the strength of the residual acceleration phase. Similarly, \citet{Zhang.Dere2006} have also reported two such phases of acceleration based on their study of 50 CMEs observed from SoHO/LASCO.

All the studies cited above use a single viewpoint to observe the CMEs. The results then inherently suffer from projection effects of the transients on to the plane of the sky. In order to overcome this, we decided to look at CMEs from the stereoscopic vision of Solar TErrestrial RElations Observatory (STEREO) \citep{Kaiser.etal2008}. The STEREO spacecraft provide two viewpoints of the prominences and the associated CMEs. We have used a stereoscopic reconstruction technique to determine the true physical coordinates of a solar feature \citep{Joshi.Srivastava2011}. The stereoscopic reconstruction would allow us to observe evolution of the true height of prominences and CMEs, and hence their true velocity and acceleration. From this we can examine if the acceleration truly exhibits bimodal profile as the model suggests. This will also give us a clue about the initiation and propagation of CMEs in the corona.

\section{Data and Observations}\label{S:observe}

\begin{figure}
\centering
\includegraphics[width=0.30\textwidth,clip=]{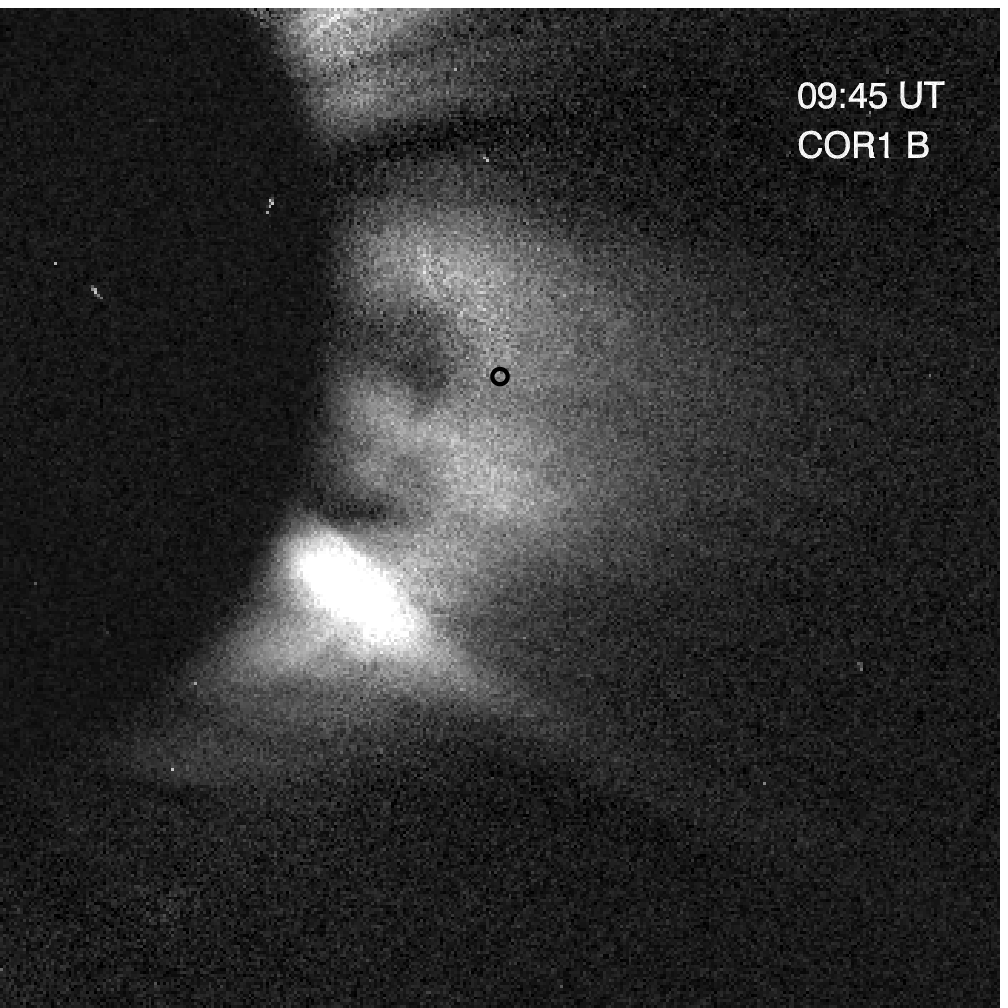}
\includegraphics[width=0.30\textwidth,clip=]{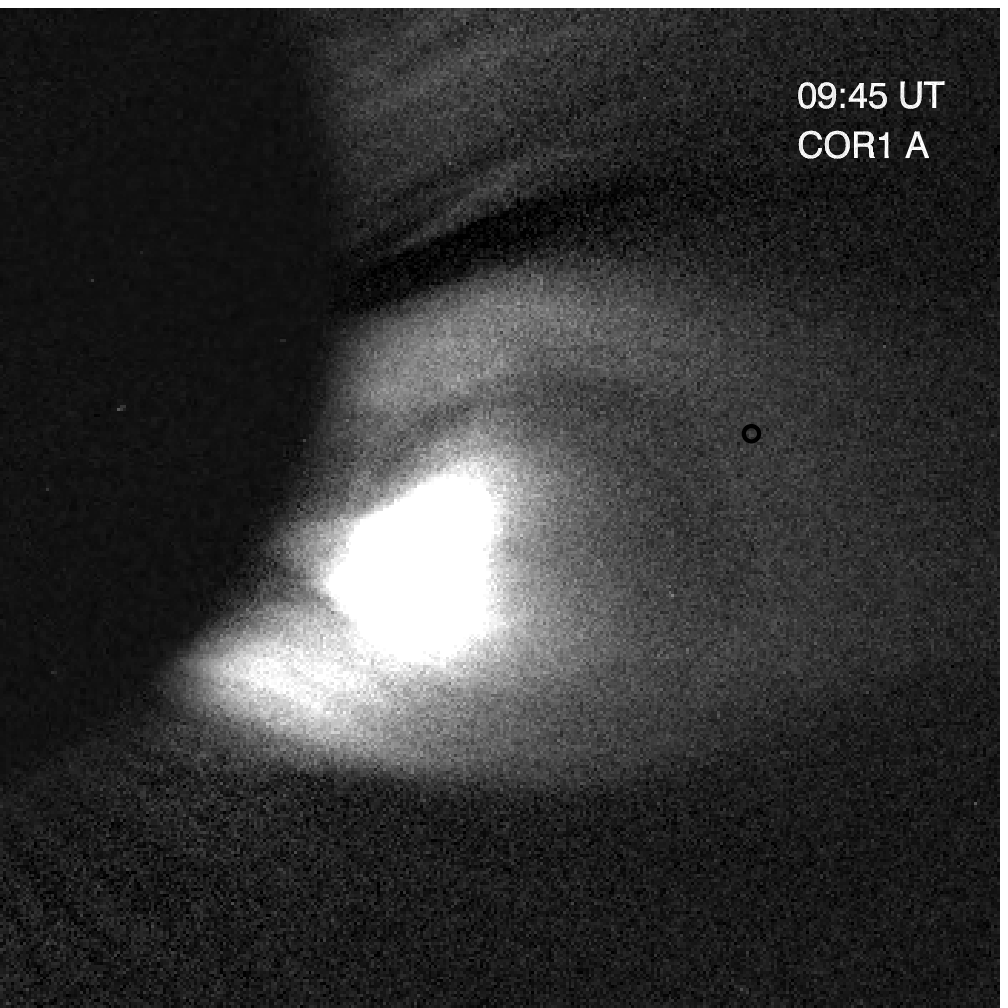}
\\
\includegraphics[width=0.30\textwidth,clip=]{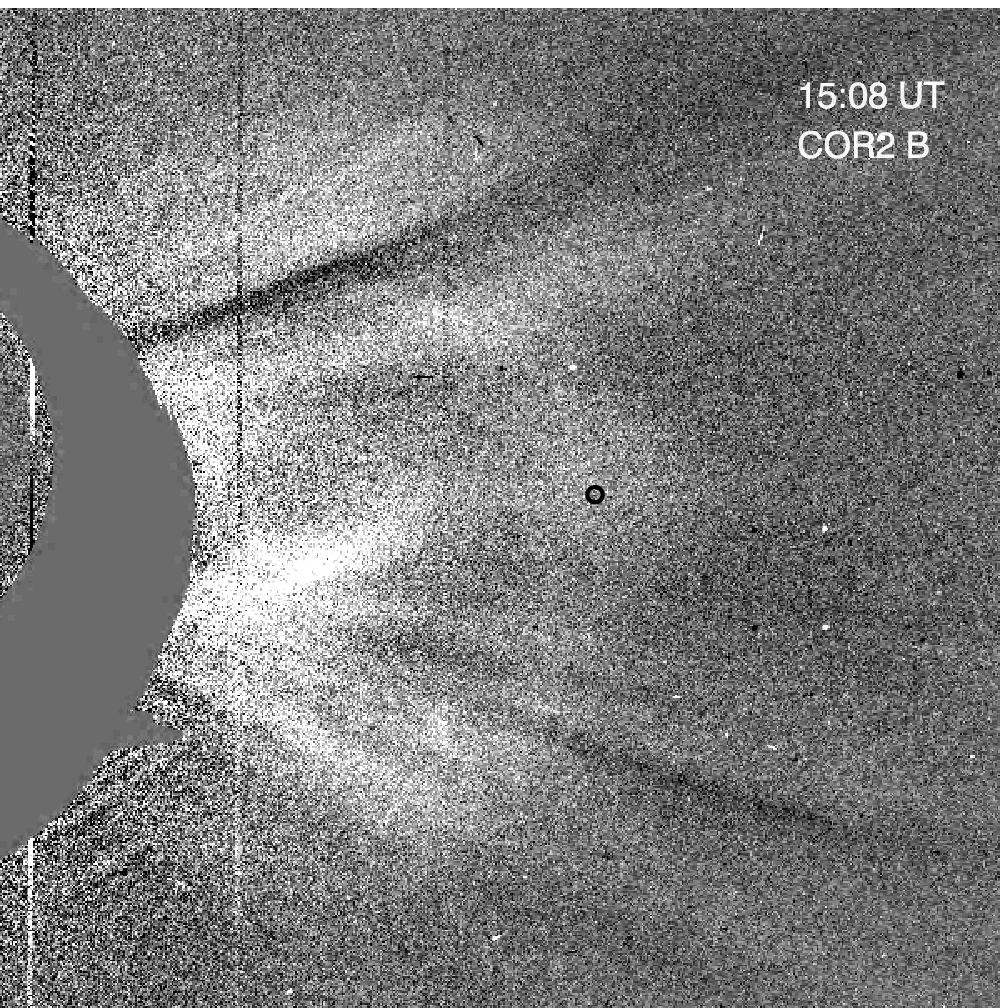}
\includegraphics[width=0.30\textwidth,clip=]{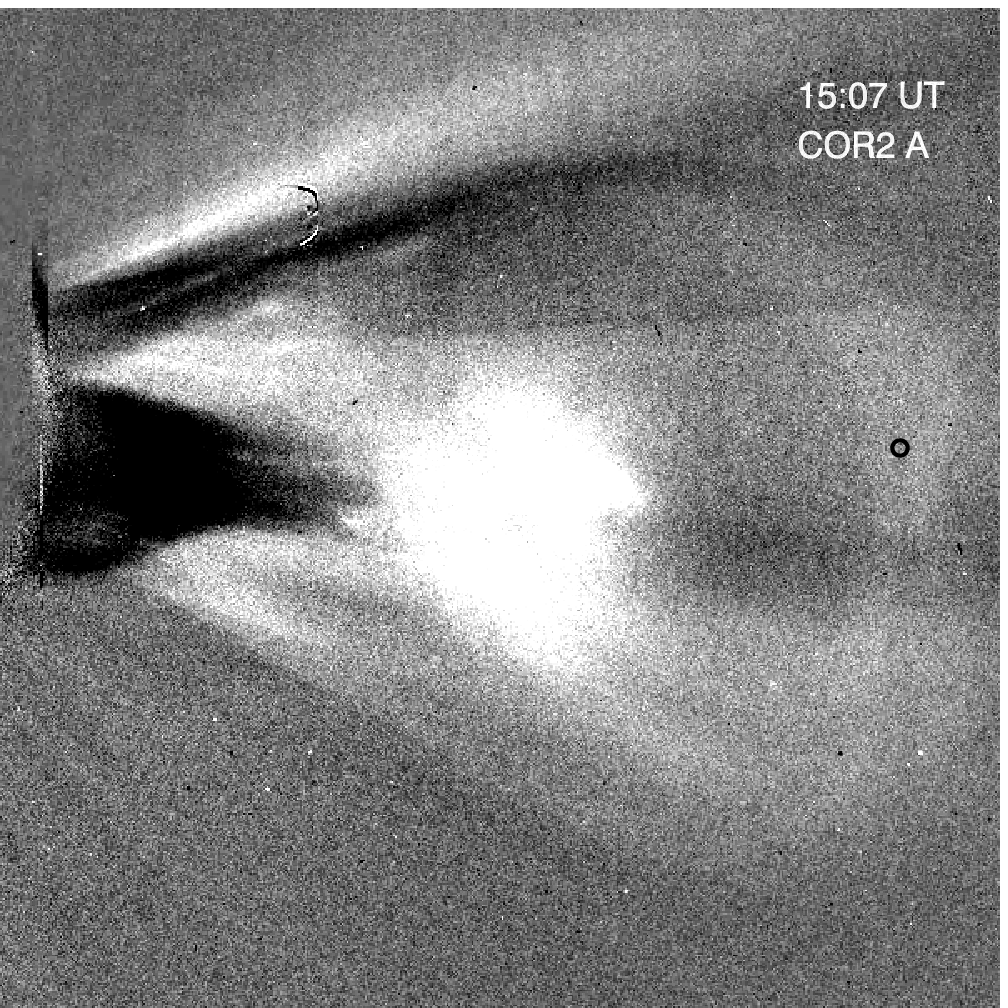}
\\
\caption{Images of the CME on 2007 November 16 seen in images from COR1 (top panels) and COR2 (bottom panels), as seen from STEREO\,B (left) and A (right).}\label{F:img16nov}
\end{figure}

\begin{figure}[!p]
\centering
\includegraphics[width=0.30\textwidth,clip=]{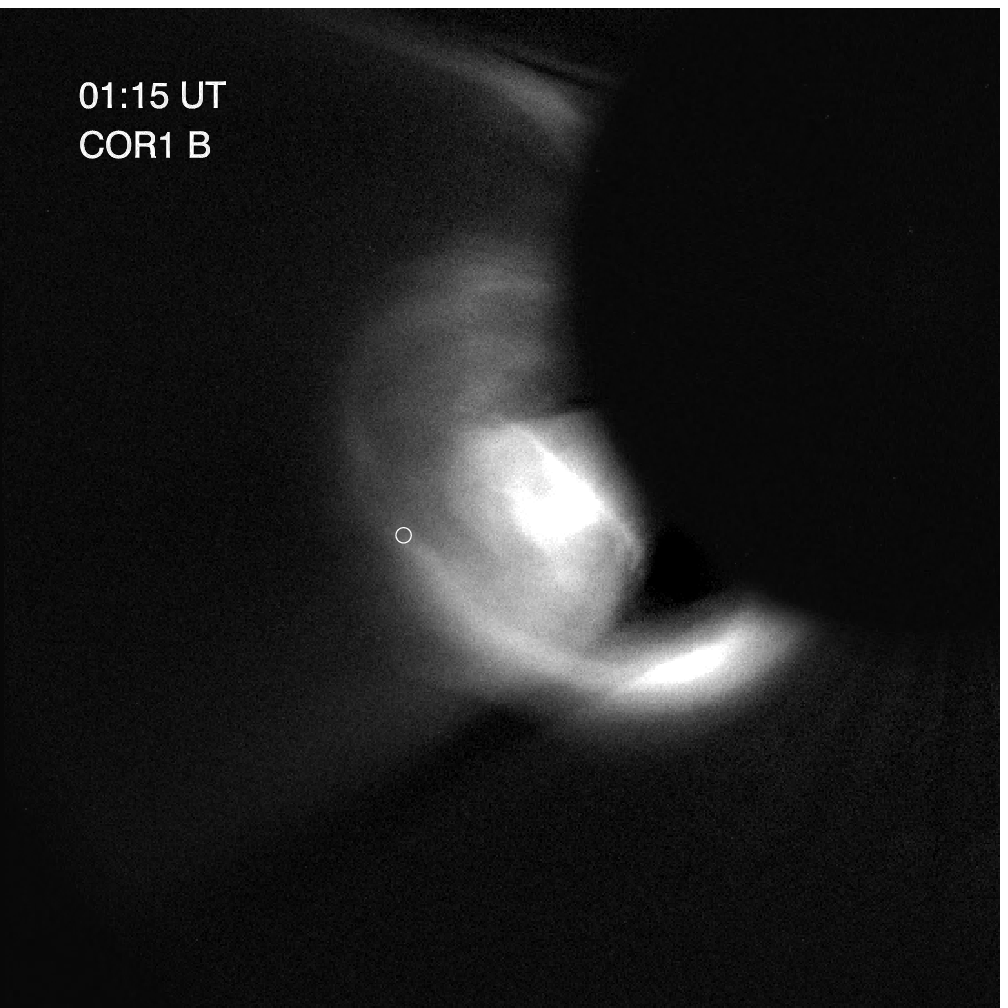}
\includegraphics[width=0.30\textwidth,clip=]{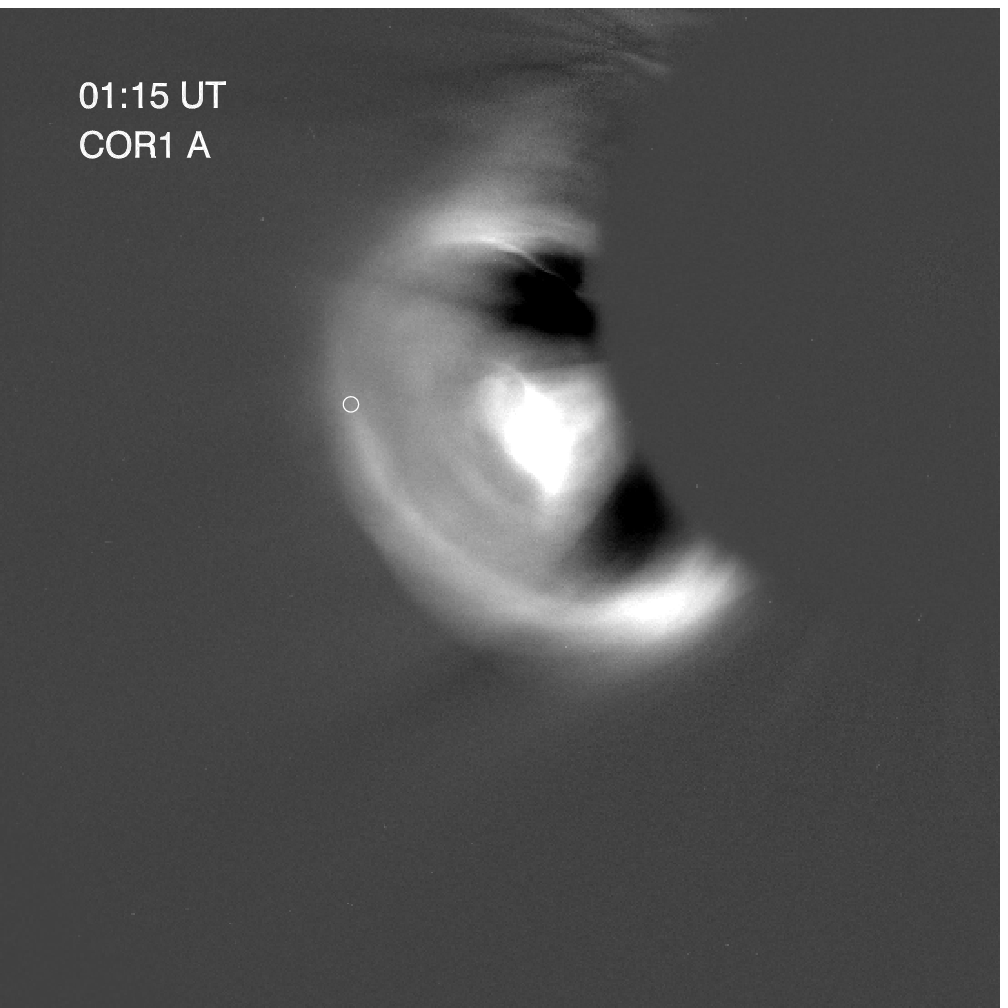}
\\
\includegraphics[width=0.30\textwidth,clip=]{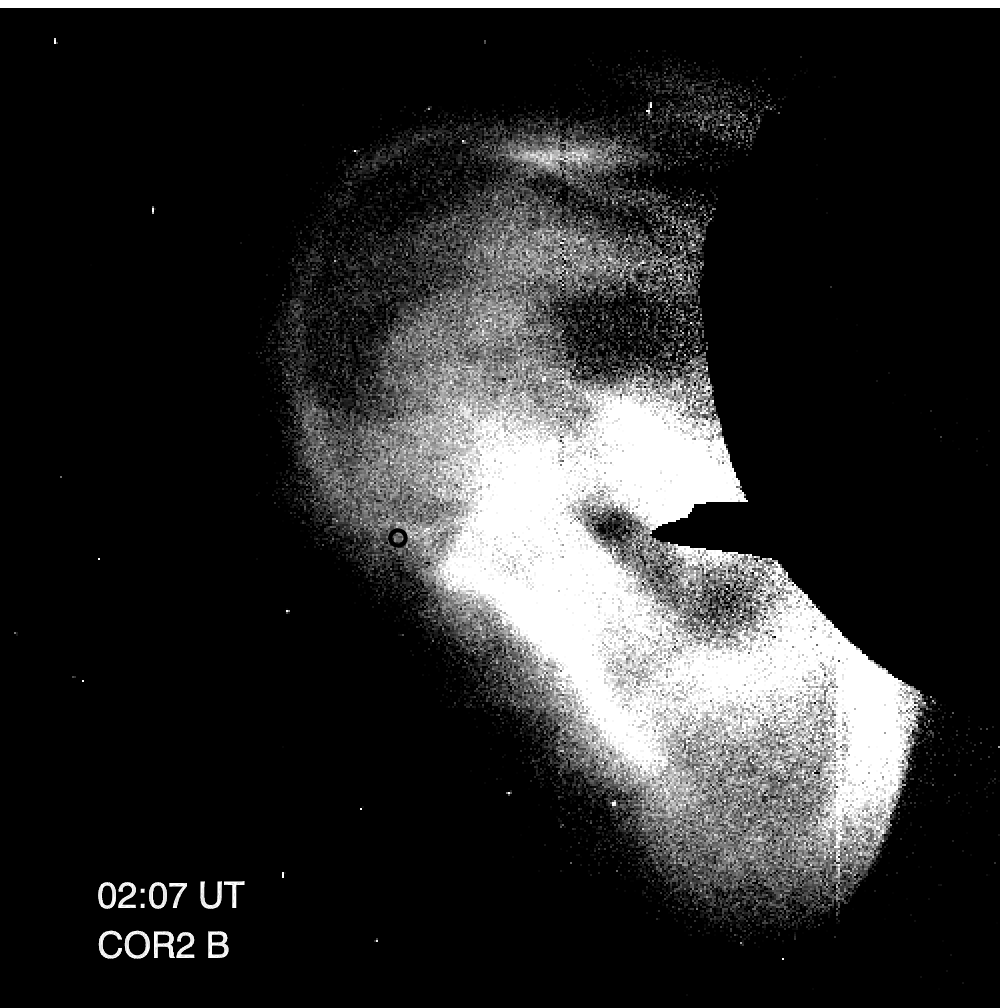}
\includegraphics[width=0.30\textwidth,clip=]{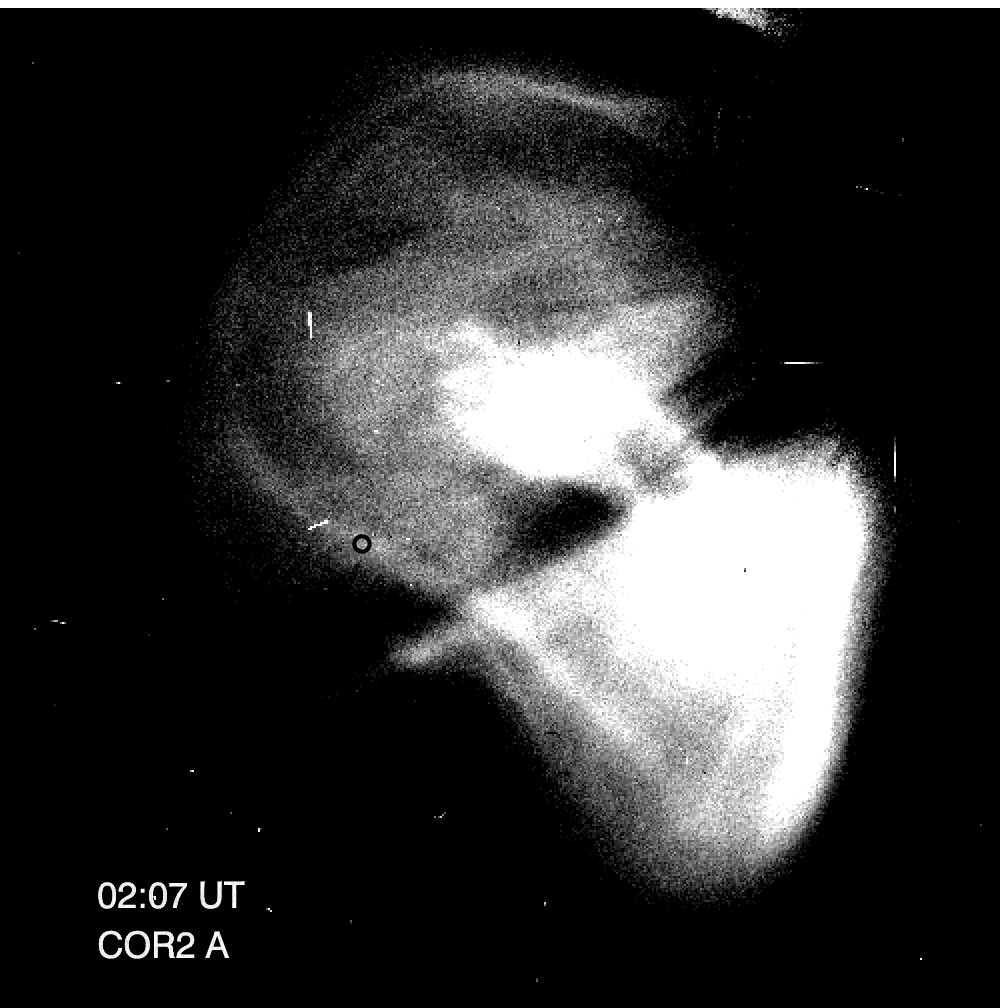}
\\
\caption{Images of the CME on 2007 December 31 seen, similar to Figure~\ref{F:img16nov}.}\label{F:img31dec}
\end{figure}

\begin{figure}
\centering
\includegraphics[width=0.30\textwidth,clip=]{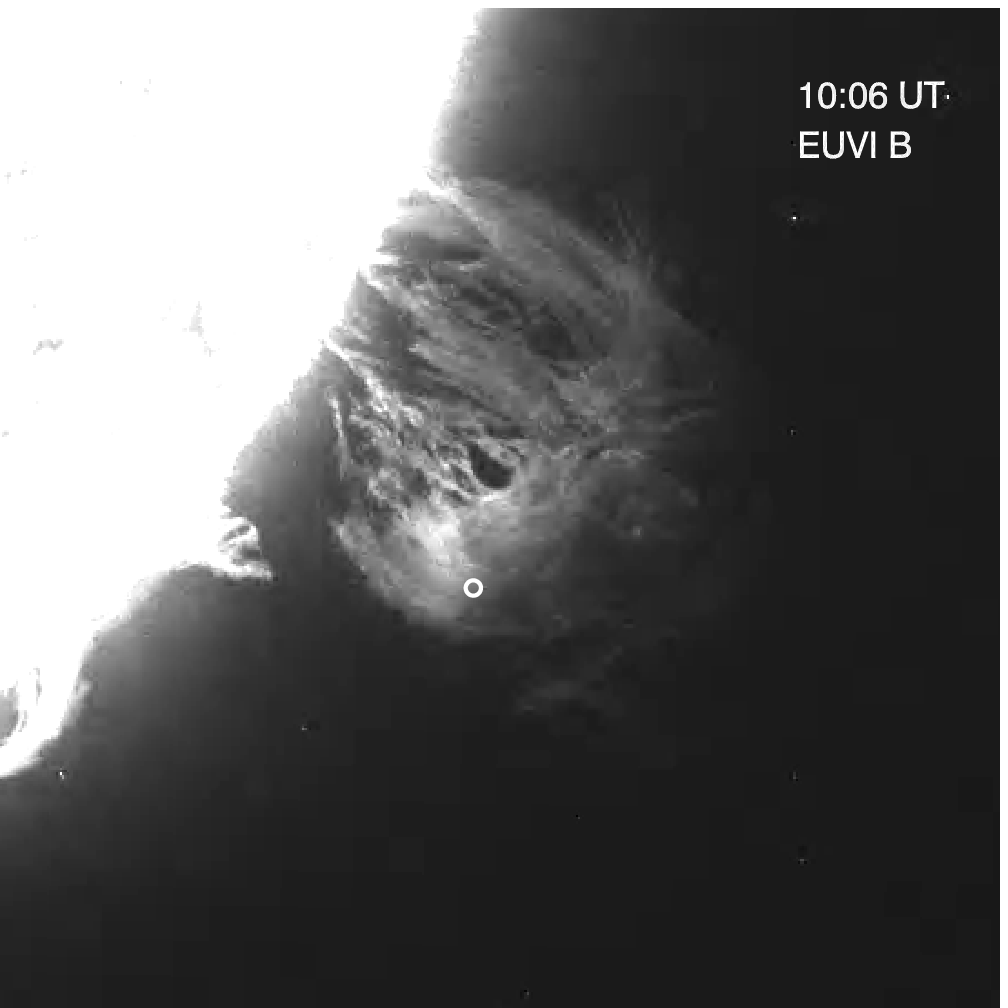}
\includegraphics[width=0.30\textwidth,clip=]{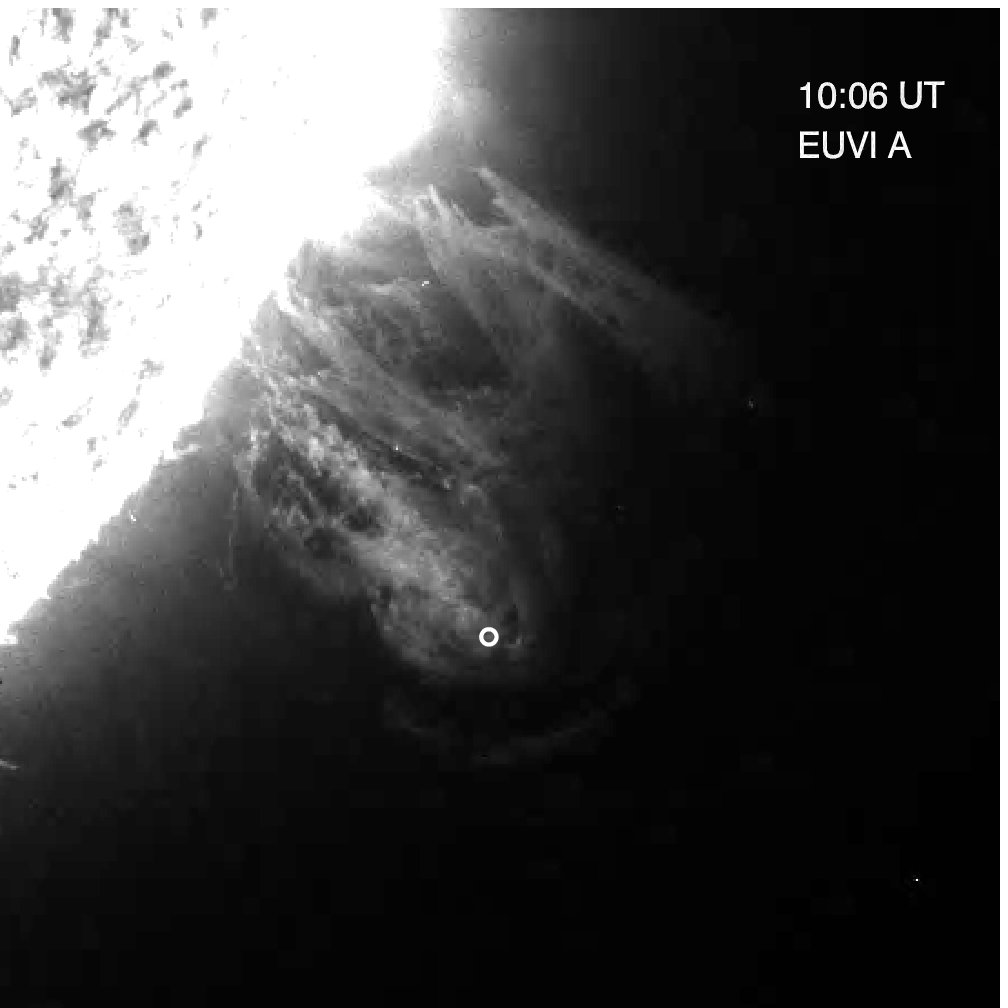}
\\
\includegraphics[width=0.30\textwidth,clip=]{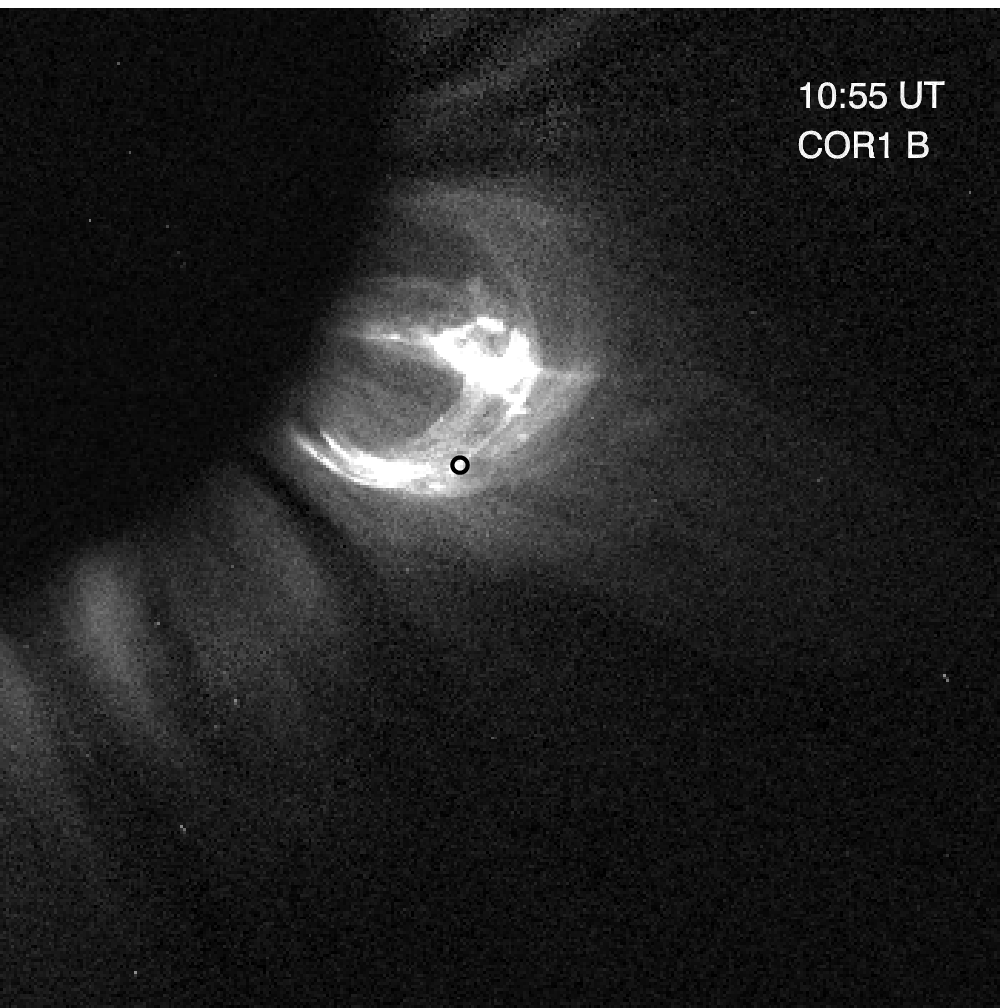}
\includegraphics[width=0.30\textwidth,clip=]{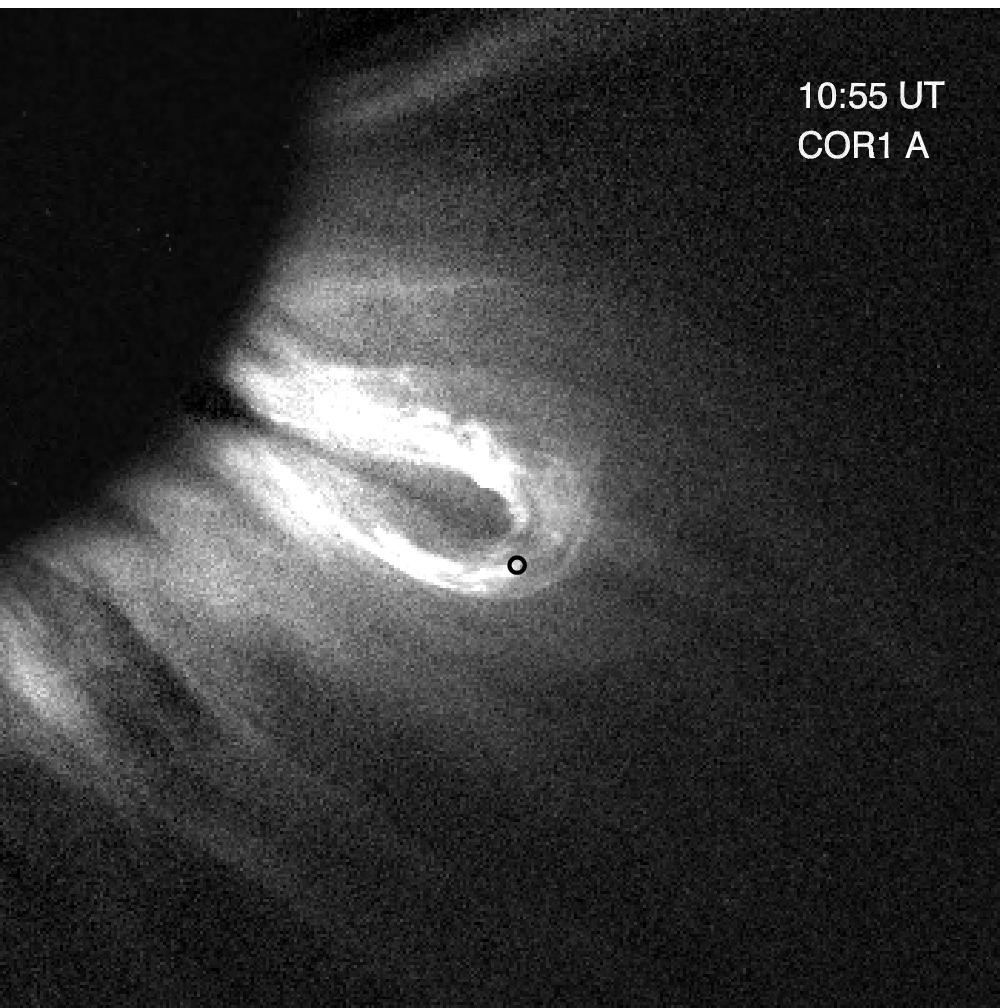}
\\
\includegraphics[width=0.30\textwidth,clip=]{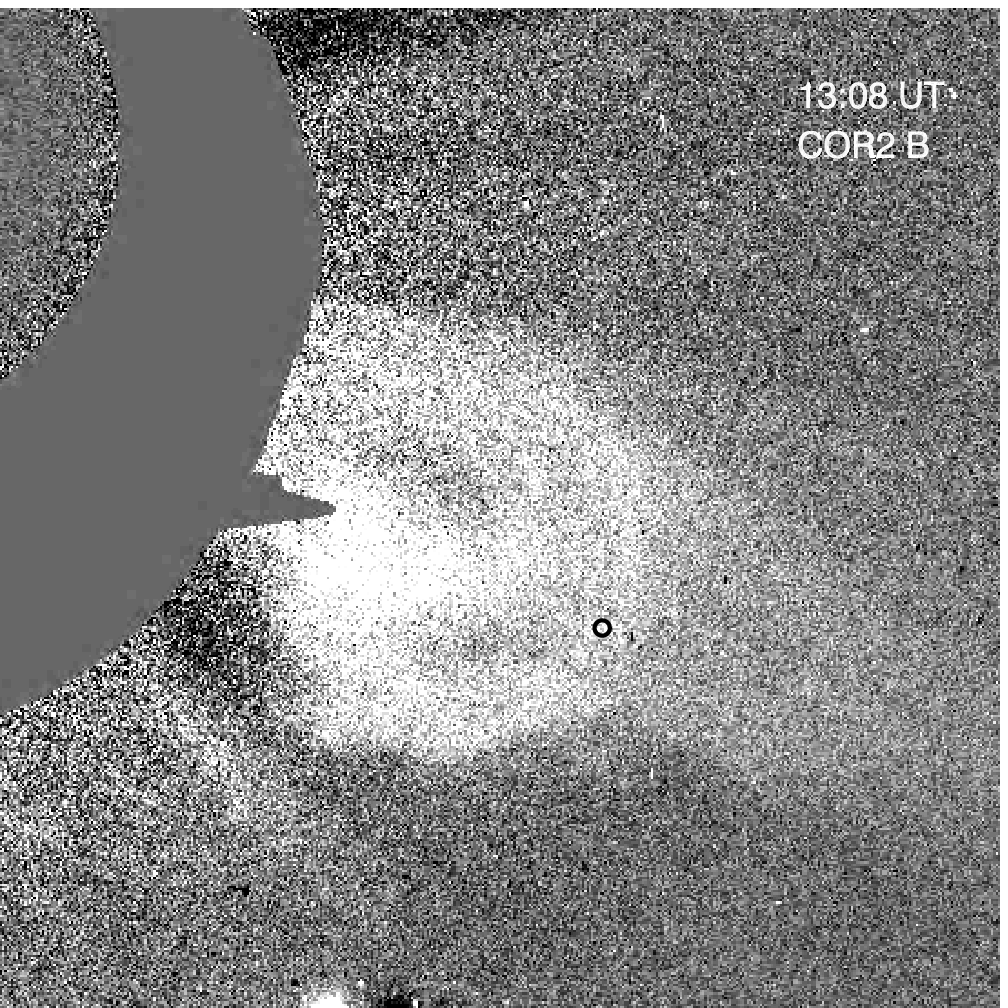}
\includegraphics[width=0.30\textwidth,clip=]{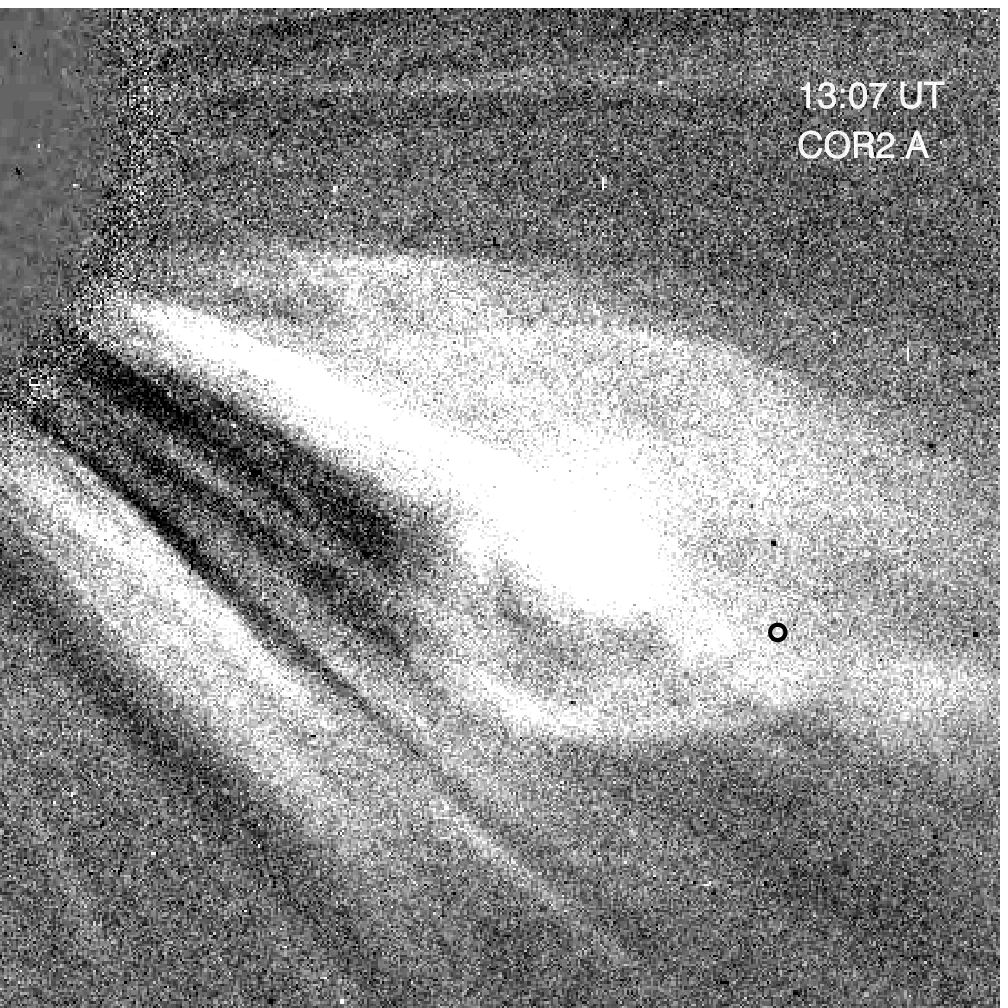}
\\
\caption{Images of the EP and associated CME on 2008 April 9, seen in images from EUVI 304\Ang\ (top panels), COR1 (middle panels) and COR2 (bottom panels), as seen from STEREO\,B (left) and A (right).}\label{F:img09apr}
\end{figure}

\begin{figure}
\centering
\includegraphics[width=0.30\textwidth,clip=]{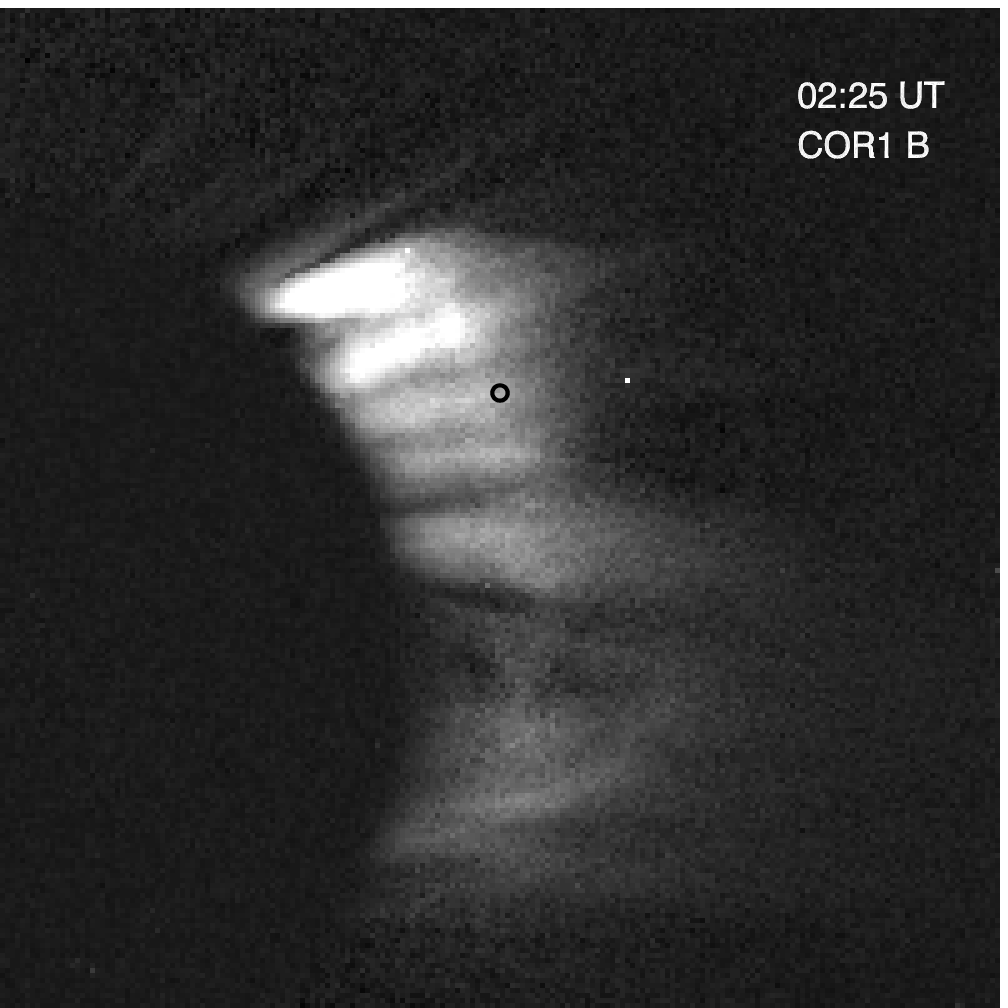}
\includegraphics[width=0.30\textwidth,clip=]{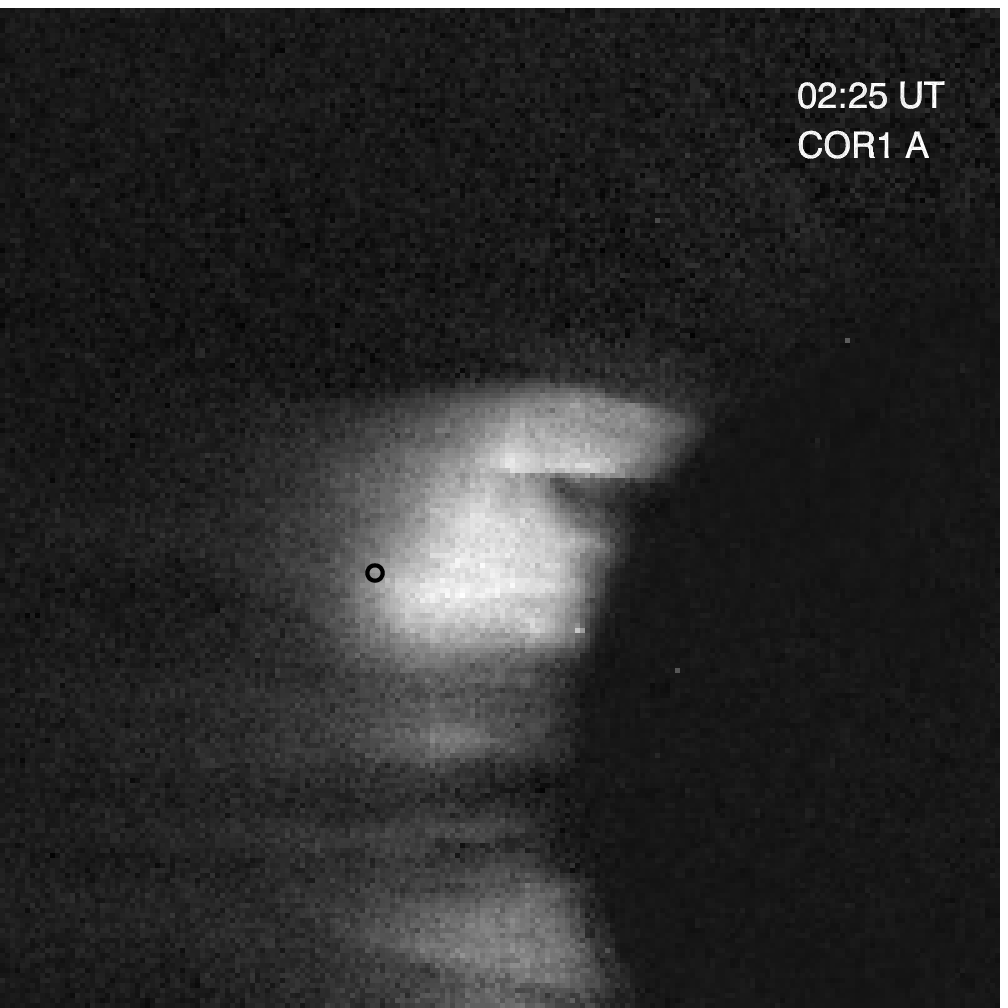}
\\
\includegraphics[width=0.30\textwidth,clip=]{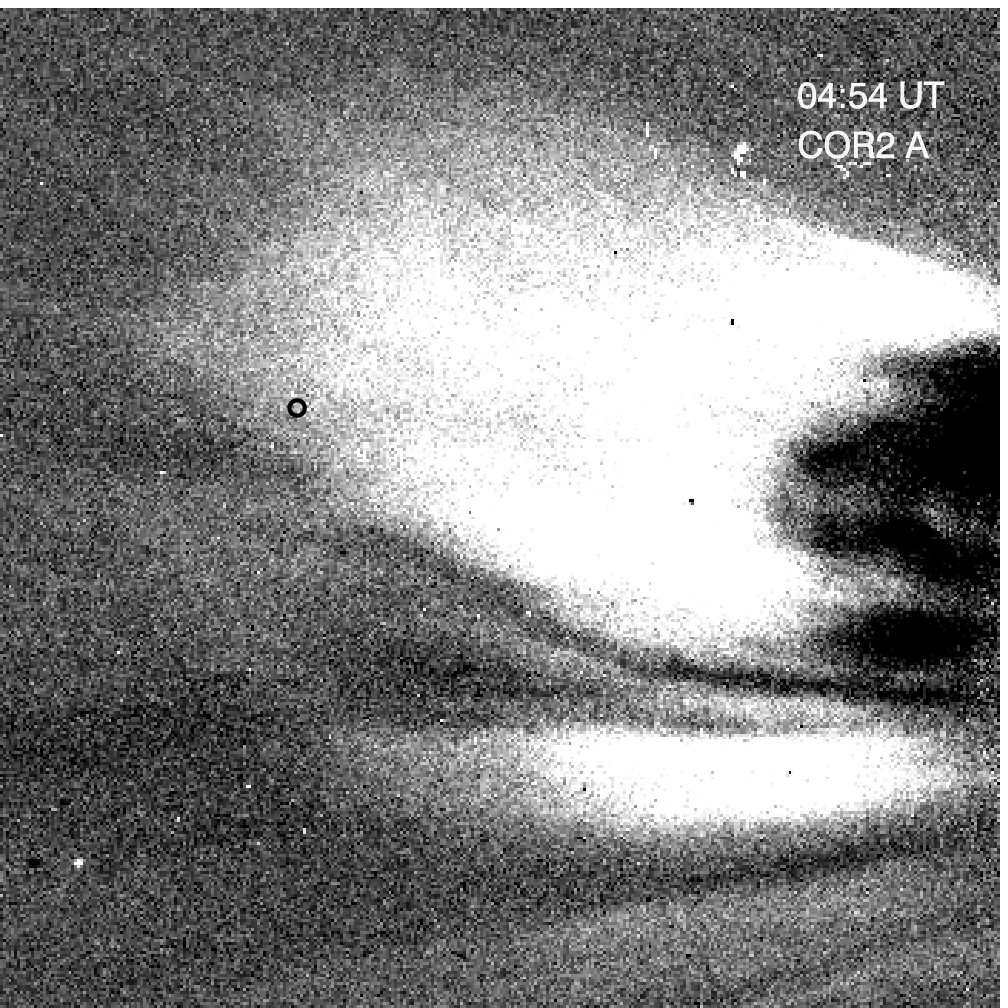}
\includegraphics[width=0.30\textwidth,clip=]{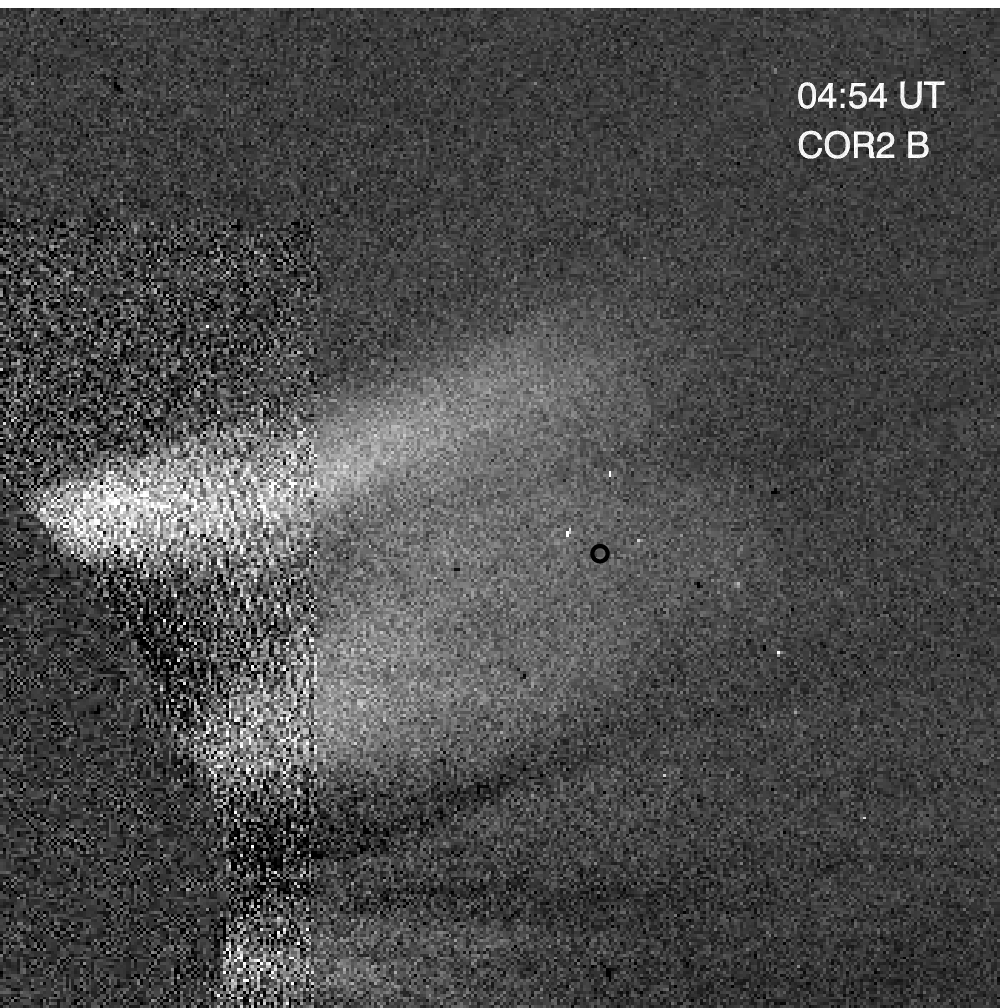}
\\
\caption{Images of the CME on 2009 December 16 seen, similar to Figure~\ref{F:img16nov}.}\label{F:img16dec}
\end{figure}

\begin{figure}
\centering
\includegraphics[width=0.30\textwidth,clip=]{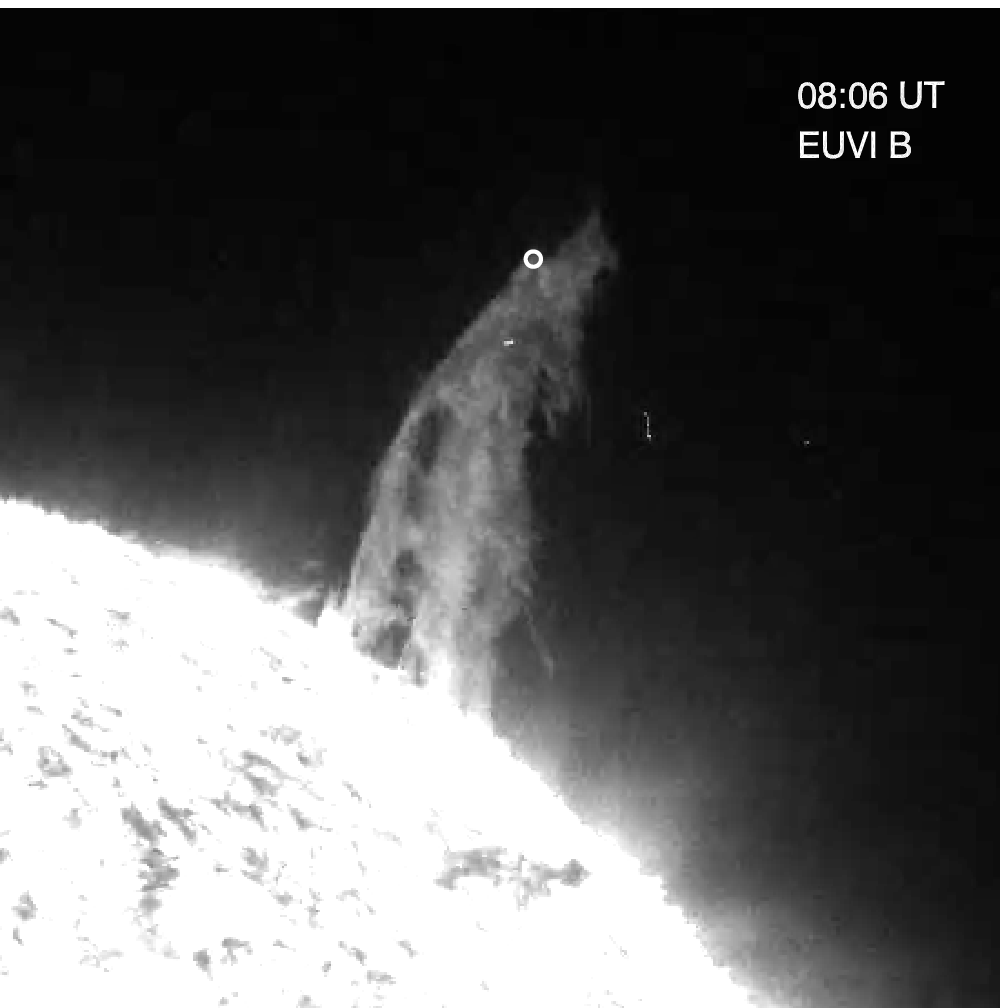}
\includegraphics[width=0.30\textwidth,clip=]{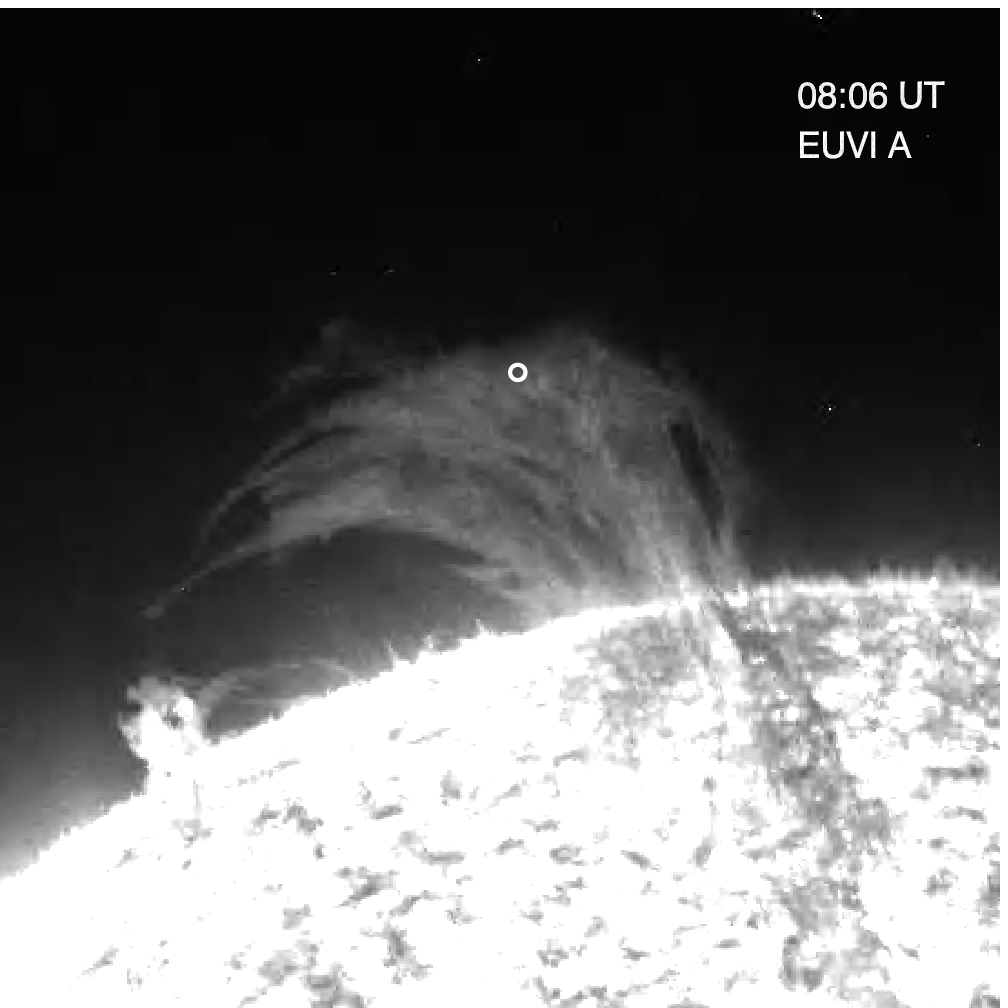}
\\
\includegraphics[width=0.30\textwidth,clip=]{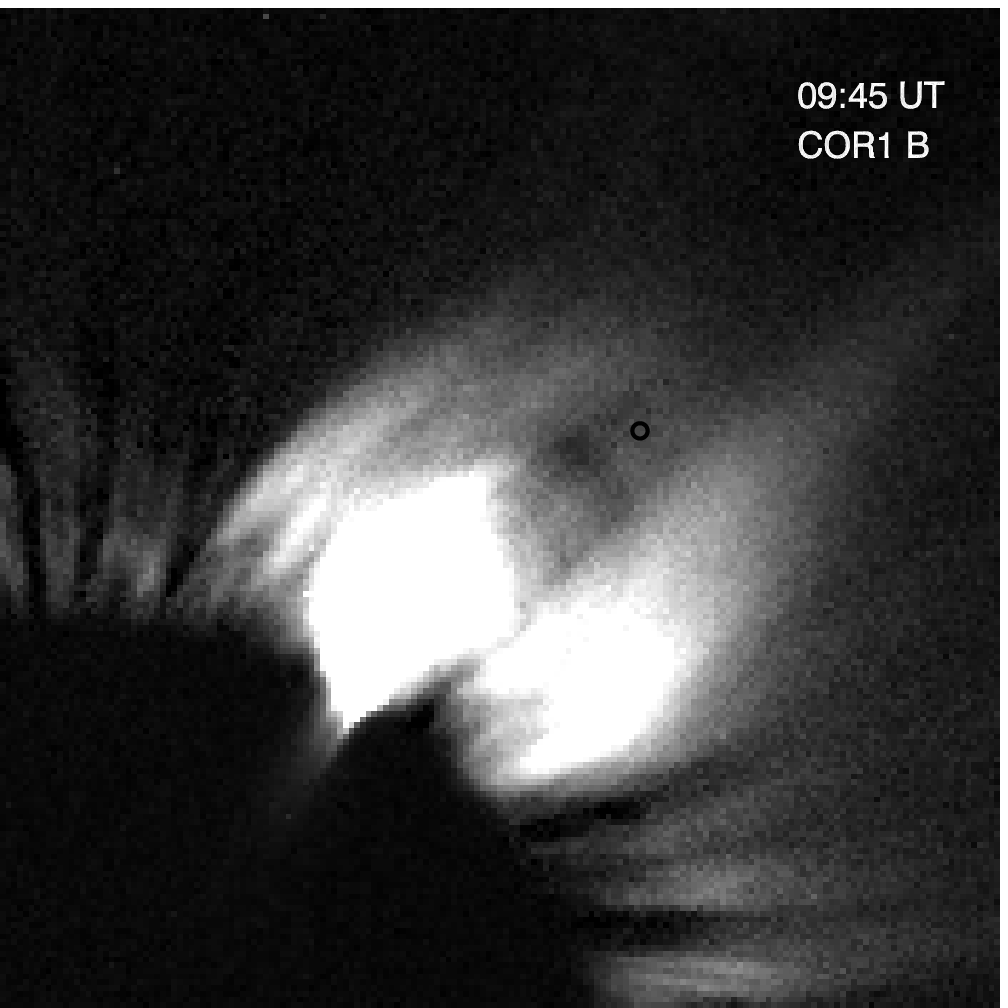}
\includegraphics[width=0.30\textwidth,clip=]{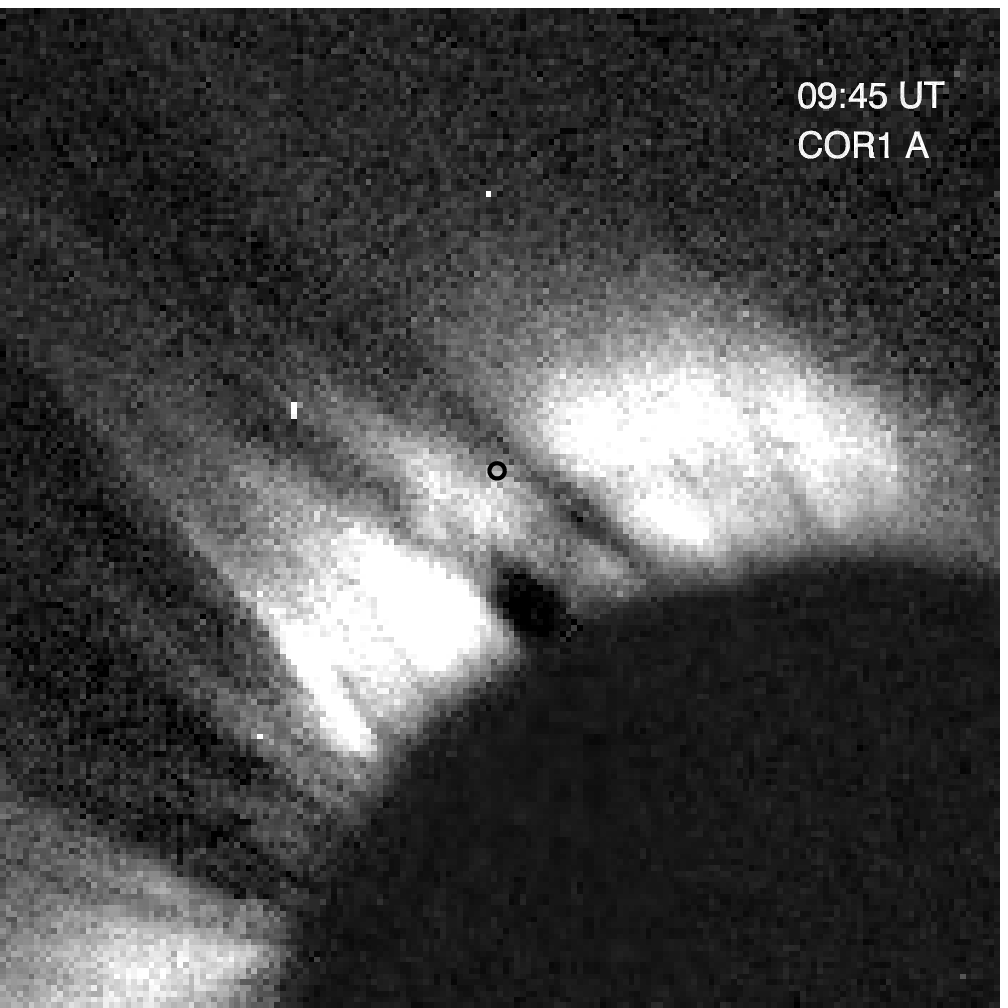}
\\
\includegraphics[width=0.30\textwidth,clip=]{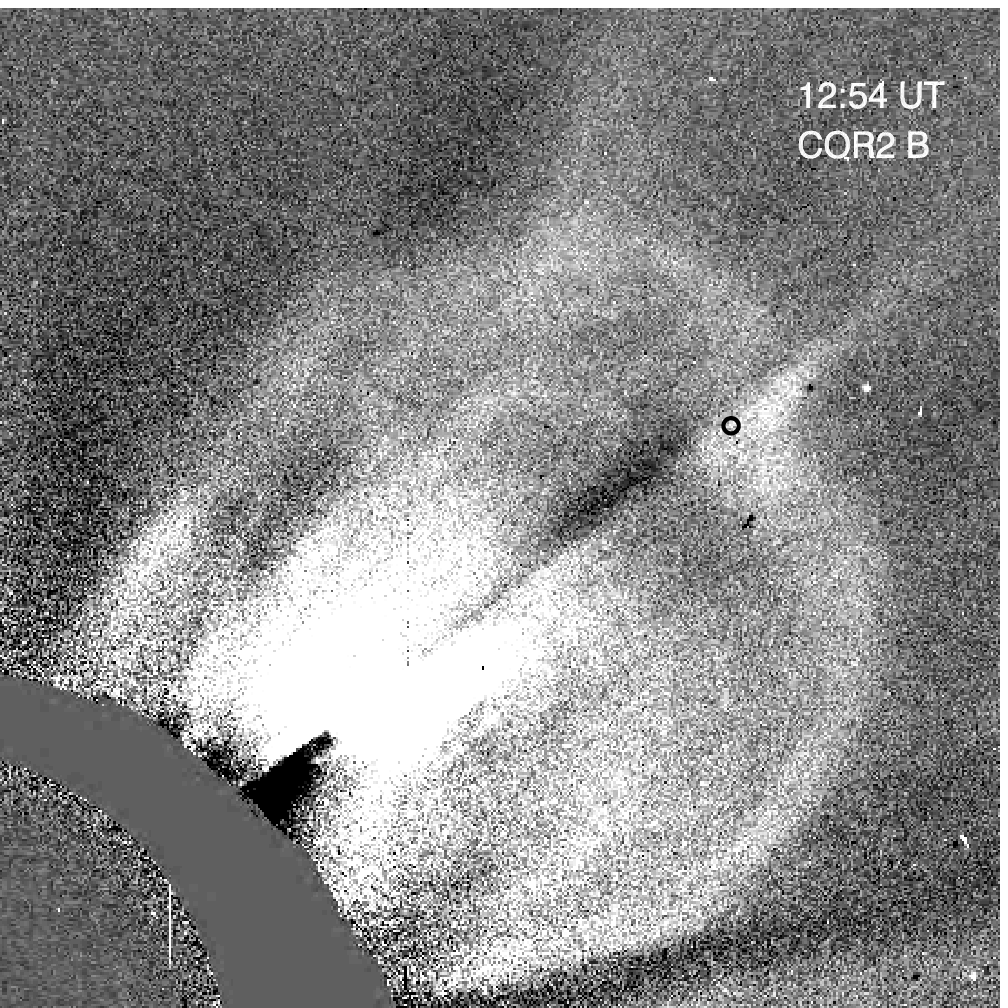}
\includegraphics[width=0.30\textwidth,clip=]{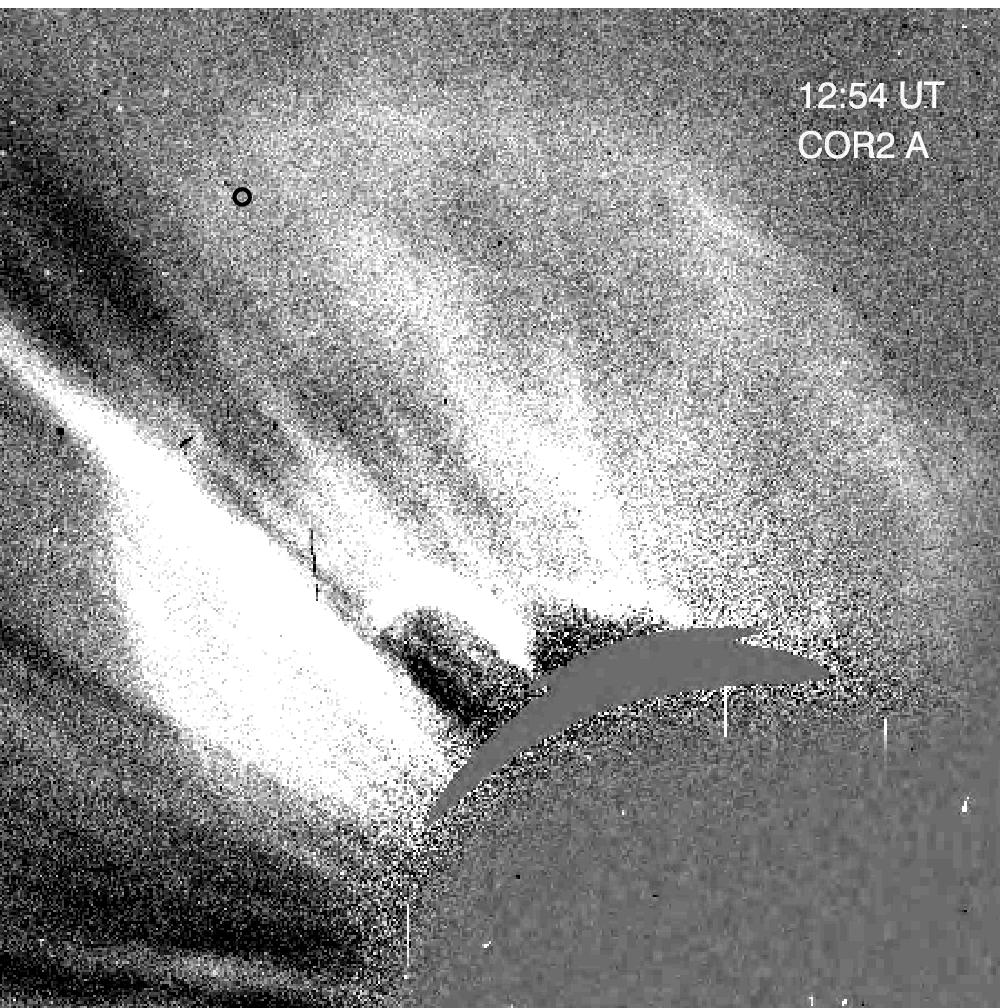}
\\
\caption{Images of the EP and the associated CME on 2010 April 13, similar to Figure~\ref{F:img09apr}.}\label{F:img13apr}
\end{figure}

\begin{figure}
\centering
\includegraphics[width=0.30\textwidth,clip=]{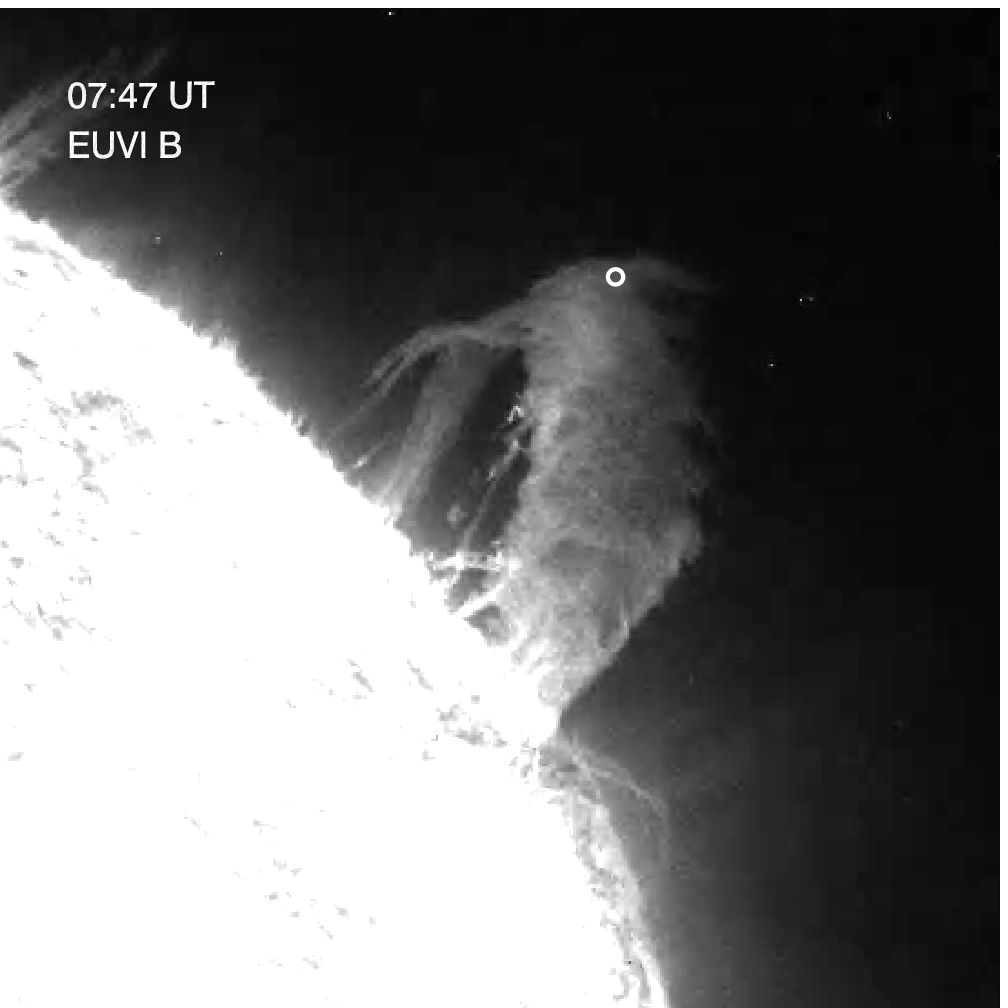}
\includegraphics[width=0.30\textwidth,clip=]{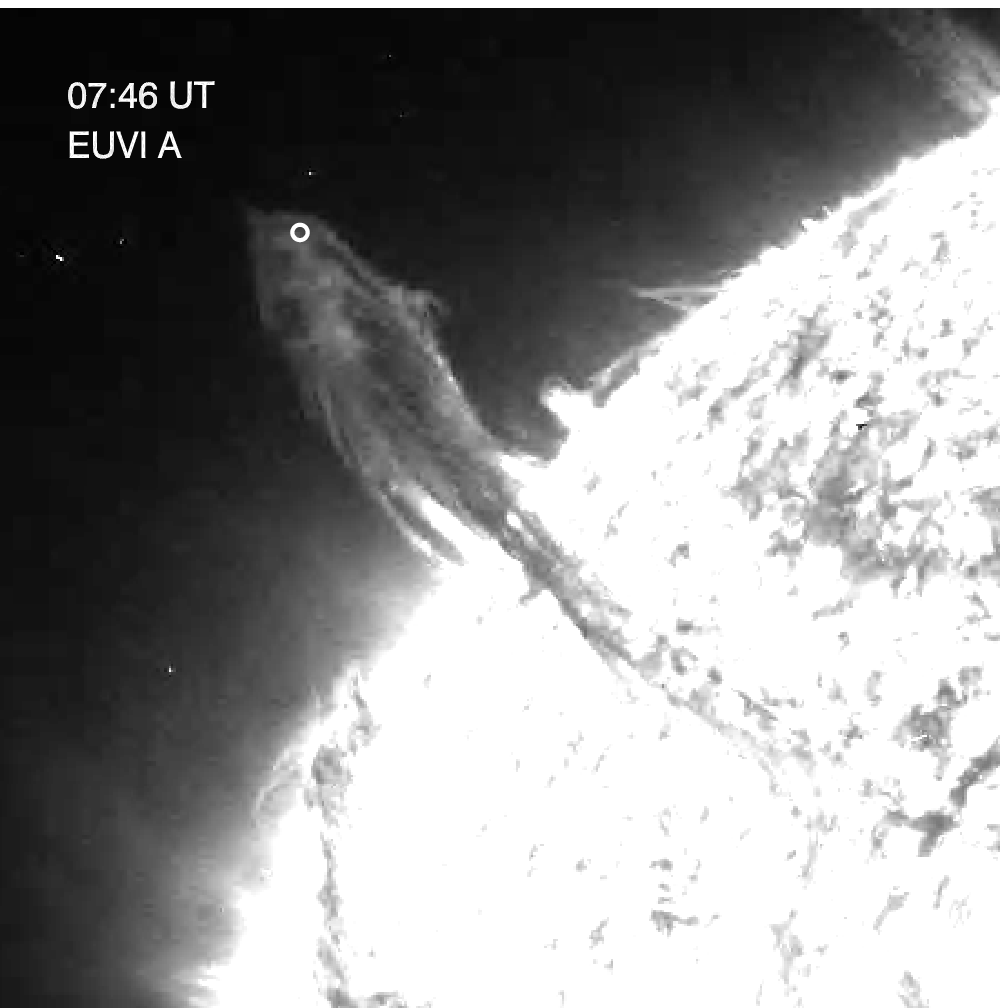}
\\
\includegraphics[width=0.30\textwidth,clip=]{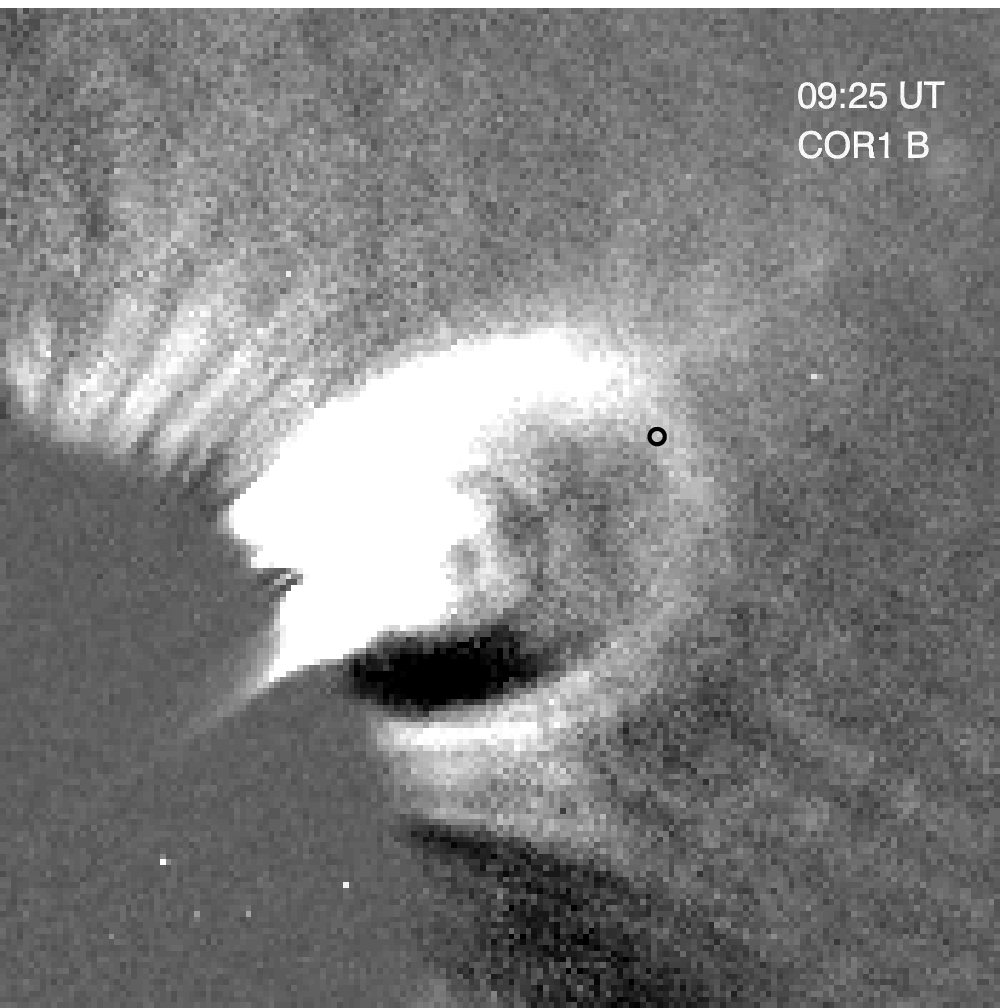}
\includegraphics[width=0.30\textwidth,clip=]{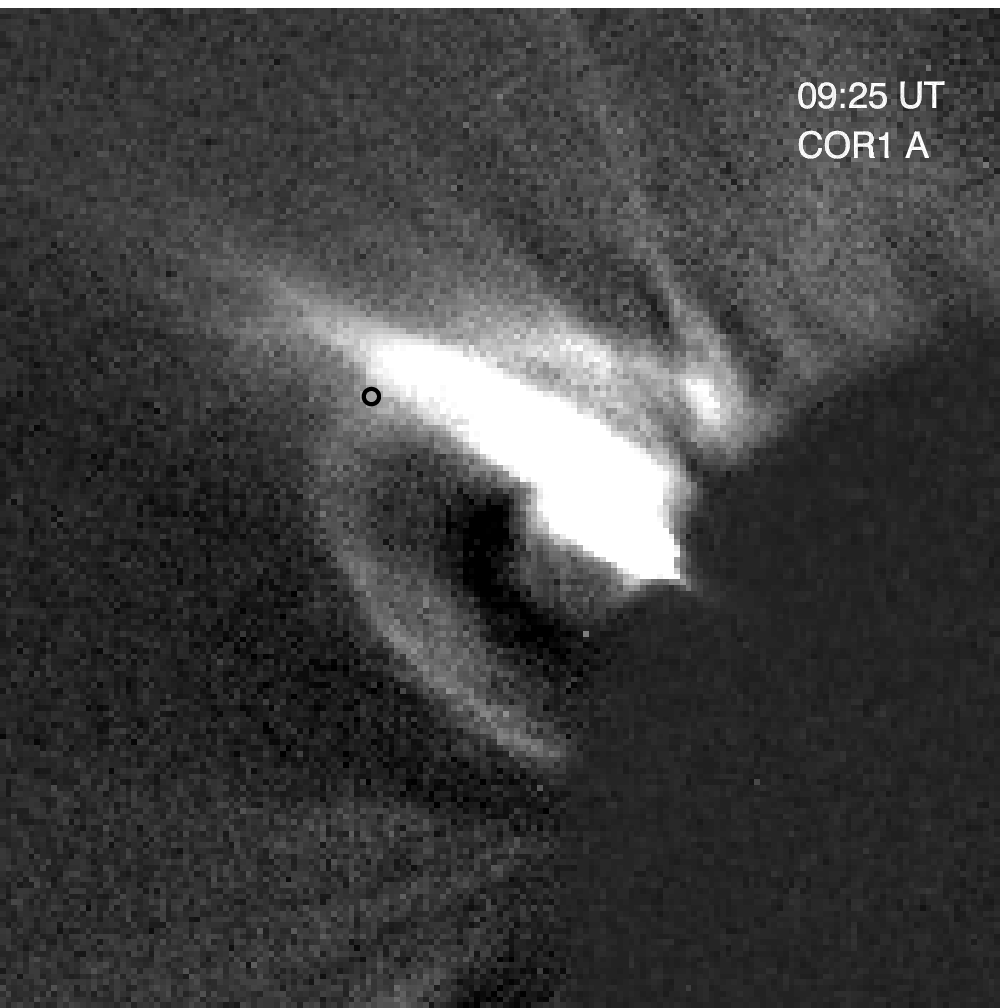}
\\
\includegraphics[width=0.30\textwidth,clip=]{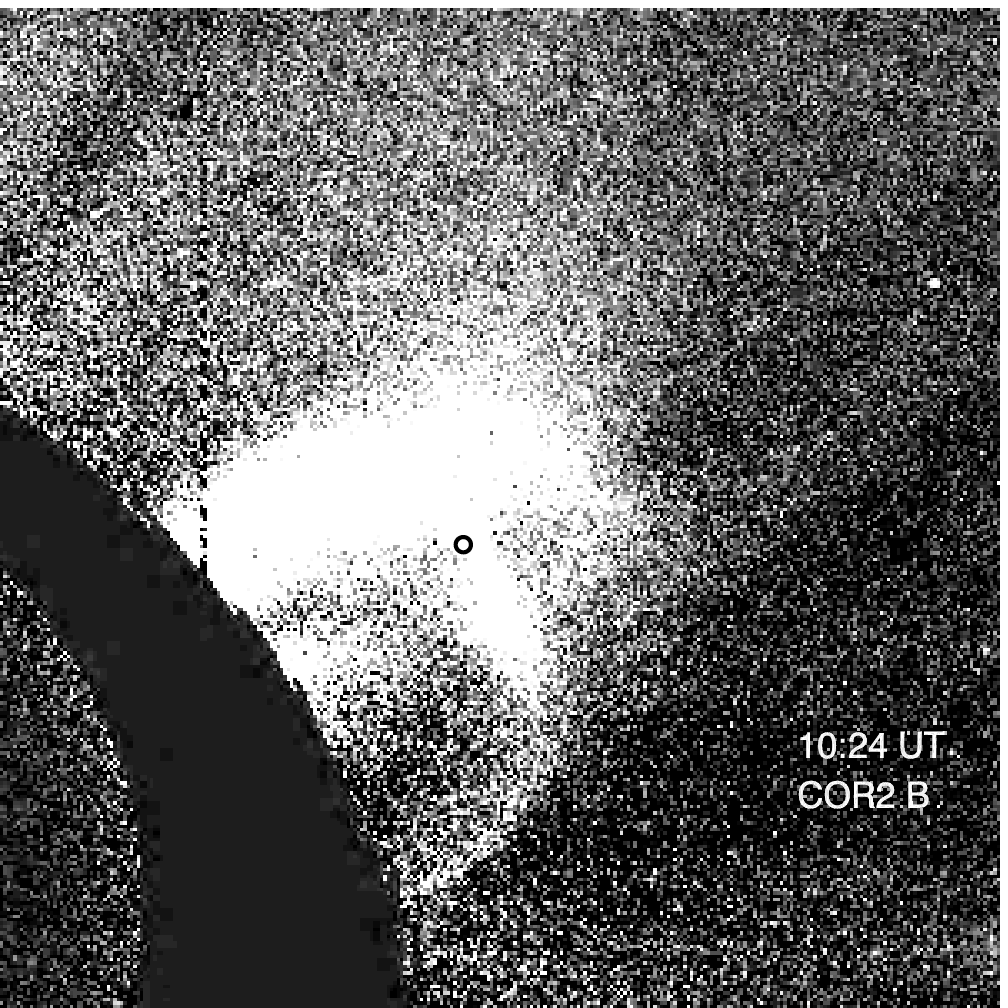}
\includegraphics[width=0.30\textwidth,clip=]{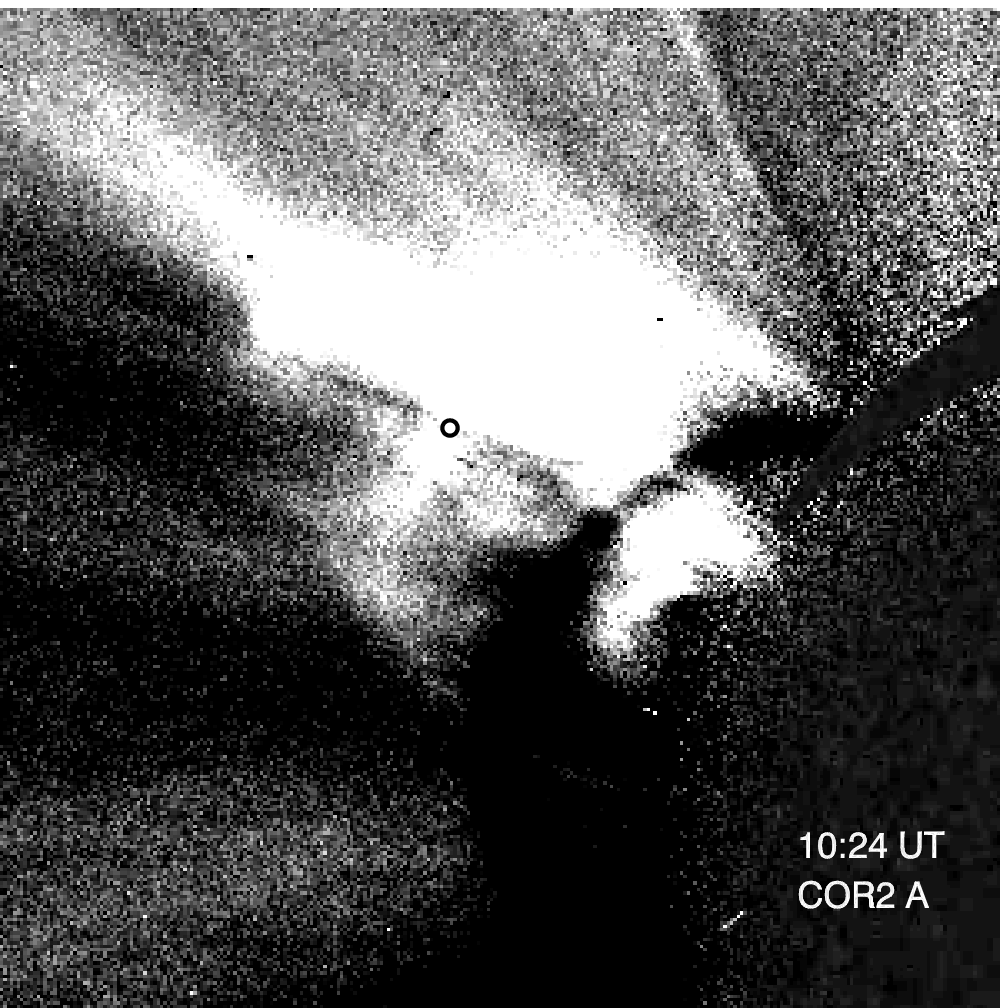}
\\
\caption{Images of the EP and the associated CME on 2010 August 1, similar to Figure~\ref{F:img09apr}.}\label{F:img01aug}
\end{figure}

The \textit{Sun Earth Connection Coronal and Heliospheric Investigation} (SECCHI) \citep{Howard.etal2008} suite of instruments on the STEREO spacecraft carries two white-light coronagraphs, COR1 and COR2, with fields of view $1.4-4.0\Rsun$ and $2.0-15.0\Rsun$ respectively, and Extreme UltraViolet Imager (EUVI) imaging the Sun at four wavelengths in the extreme ultraviolet band. We have used images obtained from the two coronagraphs to study six CMEs that occurred on 2007 November 16, 2007 December 31, 2008 April 9, 2009 December 16, 2010 April 13, and 2010 August 1. The cadence of images for the cases analysed was at best 5 minutes for COR1 and 15 minutes  for COR2. In addition, three EPs that were associated with the CMEs on 2008 April 9, 2010 April 13, and 2010 August 1 were also analysed using 304\Ang\, images from the EUVI instrument having a cadence of 10 minutes. A sample image of each event observed from coronagraphs COR1 and COR2, and EUVI 304\Ang\ for the three EPs, are shown in Figures~\ref{F:img16nov}--\ref{F:img01aug}. The soft X-ray flux data from Geostationary Operational Environmental Satellite (GOES) satellite was used to determine start and peak time of the flare associated with the CMEs on 2007 December 31 and 2009 December 16.

A feature that could be identified and tracked in all the simultaneous pairs of images from the spacecraft was used for stereoscopic reconstruction. To be able to unambiguously identify the feature in fields of view of both the STEREO coronagraphs, we had to select a feature in the inner part of the LE, and not the outermost feature in the LE. The reconstruction technique involves rotating the heliocentric Earth ecliptic coordinate system separately for STEREO Ahead (A) and Behind (B) so that one of the axes of the coordinate system lies along the Sun-spacecraft line. As a result, the plane perpendicular to this axis becomes image plane for the concerned spacecraft, and image coordinates of the feature to be reconstructed are projection of the feature in this rotated coordinate system for that spacecraft \citep{Joshi.Srivastava2011}. Using rotation matrices, and applying the epipolar constraint, it is then possible to obtain the true coordinates of the feature in heliographic coordinate system.

\subsection{Analysis}
We have used the stereoscopic reconstruction technique as described in \citep{Joshi.Srivastava2011} to obtain the true coordinates for a feature in the leading edge (LE) of all the CMEs, and the associated EPs in three of the events, in heliographic coordinate system. The reconstruction technique can be used for on-disc EUVI images, as well as coronagraph images from COR1 and COR2. The errors in determination of the height from EUVI, COR1 and COR2 are $0.02\Rsun$, $0.12\Rsun$ and $0.6\Rsun$ respectively. On fitting a polynomial function to the true height, and taking its first and second derivatives, we determine the true speed and acceleration of the EP and the LE. Since we are interested in looking at CME acceleration profiles, we only present the relevant results, and not all the information that we derive from the reconstruction, namely the Stonyhurst latitudes and longitudes. In Figures~\ref{F:res16nov}--\ref{F:res01aug}, we have shown the evolution of true height, velocity and acceleration with time. However, to know the exact height at which acceleration of the CME occurred, we have plotted the speed and acceleration as a function of the true height of the tracked feature. We have marked the reconstructed points obtained from COR1 coronagraph with plus signs, and those from COR2 with asterisks, while triangles are used to indicate features observed in EUVI 304\Ang\, images, wherever applicable. The polynomial functions used to fit the heights of the events are further used to determine errors in velocity and acceleration. The errors in heights are used in the error propagation formula, and the maximum errors in velocity and acceleration are found to be $40\kmps$ and $25\mpss$, respectively.

\section{Results and Discussion}\label{S:result}

\subsection{2007 November 16 CME}

Both the STEREO spacecraft observed this event as a faint and slow CME on the south-west limb of the Sun (Figure~\ref{F:img16nov}). The CME first appeared in the COR1\,A field of view (FOV) at 07:25\,UT, and in COR1\,B FOV at 08:15\,UT. The CME entered the COR2\,A FOV at 10:37\,UT, and in the COR2\,B FOV two hours later at 12:37\,UT. EUVI\,A 304\Ang\, images show a surge eruption on the far-side of the Sun. The surge eruption commenced at 06:26\,UT, and could be observed up to 09:46\,UT. 

Figure~\ref{F:res16nov} shows results of the stereoscopic reconstruction technique applied to a feature on the LE of this CME. The plots on the left in Figure~\ref{F:res16nov} show change in height, and the resultant speed and acceleration obtained from derivatives of a polynomial function fitted to the true height, as a function of time. The plots on the right show speed and acceleration as a function of true height. From this figure we find that the CME speed increases very rapidly in the COR1 FOV, and is almost constant in COR2 FOV. The acceleration of the CME in the COR1 FOV is $50\mpss$, but it falls rapidly, and is down to $11\mpss$ at about $3.7\Rsun$. Thus, maximum value of acceleration, and the height at which the CME attained this value, are not available to us. We however point out that the maximum acceleration of the CME occurred at or less than $2\Rsun$ in height. In the higher corona, \thatis, in the COR2 FOV we see that the acceleration once again rises by a small amount to reach $18\mpss$ at a height of around $8\Rsun$, which can be attributed to the residual acceleration phase consistent with the model proposed by \citet{Chen.Krall2003}.

\subsection{2007 December 31 CME}

This was a bright CME with a well-defined symmetrical LE on the south-east limb of the Sun, as seen in Figure~\ref{F:img31dec}. The CME was associated with a C8 class flare which originated in NOAA active region (AR) 10980. The CME appeared in the COR1\,B FOV at 00:55\,UT, and at 01:00\,UT in COR1\,A FOV. The CME crossed the COR1 FOV in about one hour, indicating that it was a relatively fast CME, and appeared in the COR2 FOV at 01:37\,UT in the two spacecraft. This CME showed an unusual cusp in its LE, which was distinctly visible in the COR2 images. \citet{Thernisien.etal2009} have used the graduated cylindrical shell model \citep{Thernisien.etal2006} to fit two shells flanking the cusp for this CME, and employed the forward modelling technique to determine its true direction of propagation and speed. We have used this feature for the purpose of reconstruction. On the eastern limb of the 304\Ang\, image from EUVI\,B, a flare can be seen at 00:46\,UT, followed by opening up of the field lines which can be clearly seen in 171\Ang\, and 195\Ang\, images from EUVI\,B. From the GOES soft X-ray flux data, we find that the flare started at 00:45\,UT, and peaked at 01:03\,UT.

The CME speed increased in the lower corona to reach $812\kmps$ at a height of $2.8\Rsun$, and showed a little dip before attaining a constant value of around $870\kmps$ (Figure~\ref{F:res31dec}). This CME has the highest value of maximum acceleration of all the CMEs studied here, which is over $1500\mpss$ at a height of about $2\Rsun$. Such a high value is observed for CMEs associated with flares, which are termed as impulsive by \citet{Sheeley.etal1999} and \citet{Moon.etal2002}. However, the flare in this case was classified with X-ray class C8, and such high values of CME acceleration are earlier reported to be associated with X-class flares. Such a high value of acceleration has also been reported by \citet{Alexander.etal2002} for a CME associated with an X1.2 class flare. Also, assuming that this CME achieved its maximum value somewhere below $2\Rsun$, we note that this favours the CME model proposed by \citet{Chen.Krall2003}, where they predict a bimodal acceleration profile. This CME has been previously been analysed by several researchers. Among them, \citet{Temmer.etal2010} have found acceleration of this CME to be $1300\mpss$, while \citet{Lin.etal2010} have found it to be over $1000\mpss$; both the results obtained from stereoscopic reconstruction of the CME. The large difference in acceleration values between our results, and those cited above, can be attributed to the different assumptions involved in the numerous reconstruction techniques \citep{Mierla.etal2010}.

\subsection{2008 April 9 CME and EP}

The CME on 2008 April 9 was associated with an active-region prominence, and was observed on the south-west solar limb, as shown in Figure~\ref{F:img09apr}. The CME first appeared in the COR1\,A and B FOVs at 10:15\,UT and 10:25\,UT respectively, while it could be just seen in COR2\,A and B FOVs at 12:07\,UT. The LE showed a bright knot close to its highest point, which was tracked during the reconstruction. The prominence material could be seen in 304\Ang\, images from 09:26 onwards in EUVI\,A and 09:46\,UT onwards in EUVI\,B.

This CME LE showed very smooth changes in both its speed and acceleration, as can be seen from Figure~\ref{F:res09apr}. The speed increased till the CME reached about $4\Rsun$, but the peak of the acceleration profile could not be observed, which, as in previous cases, occurred at a height less than $2\Rsun$. The acceleration kept on decreasing till it reached a value of around $-14\mpss$. Correspondingly, the maximum speed was $530\kmps$, and it decreased till the time the LE reached a height of $\sim11.5\Rsun$. The prominence associated with this CME showed an increase in acceleration up to a height of almost $4\Rsun$, until the time it could be observed in COR1 images. However, velocity of the prominence ($\sim250\kmps$) was found to be less than that of the CME ($>400\kmps$).

\subsection{2009 December 16 CME}

This CME was associated with a C5.3 class flare in NOAA AR 11035 that started at 00:59\,UT, and had its peak at 01:09\,UT. The CME first appeared in the COR1\,A and B FOVs at 01:40\,UT and 01:35\,UT respectively, while it could be just seen in COR2\,A and B FOVs at 03:08\,UT and 03:39\,UT respectively (Figure~\ref{F:img16dec}). 

From the COR1 observations, we could not see the peak of acceleration of the CME LE, we could only observe the acceleration decrease from around $90\mpss$ to $0\mpss$ as the CME travels from a height of $2\Rsun$ to $6\Rsun$, as seen in Figure~\ref{F:res16dec}. At this height, the CME velocity attains a value of $350\kmps$. Later, at a height of around $11\Rsun$, we find that acceleration shows a smaller rise before reaching a value of $-20\mpss$ at $14\Rsun$.

\subsection{2010 April 13 CME and EP}

This CME was associated with a large northern polar crown filament, as seen in Figure~\ref{F:img13apr}. The filament appeared edge-on on the western limb in the EUVI\,B FOV, but it could be seen extending from the central meridian right up to the north-east limb in EUVI\,A. The prominence eruption commenced at 08:36\,UT, and the CME LE could be seen at 08:50 and 08:40\,UT respectively in COR1\,A and B FOVs. The CME LE could be seen in the COR2 FOV at 10:39\,UT. The prominence material could also be very conspicuously seen in images from both the coronagraphs on the two spacecraft.

This CME showed changes in speed and acceleration similar to the one on 2007 November 16 (Figures~\ref{F:res13apr} and \ref{F:res16nov}). Its speed very rapidly reached a value of $300\kmps$ at height $3.8\Rsun$, and during the same time its acceleration dropped from $60\mpss$ to $27\mpss$. The peak of the acceleration however could not be observed. Like the event of 2008 April 9, here too we find the prominence showing an increasing acceleration at least till $4\Rsun$.

\subsection{2010 August 1 CME and EP}

This CME was also associated with a northern polar crown filament, as seen in Figure~\ref{F:img01aug}. The filament appeared as a hedgerow prominence in EUVI\,B 304\Ang\, images, while the line-of-sight was along the spine in EUVI\,A images. The CME was first seen in COR1\,A FOV at 08:10\,UT, and at 08:25\,UT in COR1\,B FOV. Due to a data gap in the COR2 observations, the CME was seen only in a single image at 10:24\,UT in COR2 A and B.

This CME behaved very differently from the rest analysed in this study. Its speed was very low at the start, and it gradually reached the maximum speed of $567\kmps$, at a height of $\sim4.5\Rsun$. At this height, the CME was still accelerating, but owing to a data gap in the COR2 observations, its peak value could not be determined. The maximum acceleration of the prominence was $40\mpss$ at a height of about $1.5\Rsun$, and then showed a steady decrease to around $10\mpss$ at a height of $3\Rsun$. The CME LE accelerated very late into its eruption. At a height of $4\Rsun$, the LE showed an acceleration value of over $200\mpss$.

\section{Summary and Conclusions}\label{S:conclude}

\begin{figure}
\centerline
		{
	\includegraphics[width=0.465\textwidth,clip=]{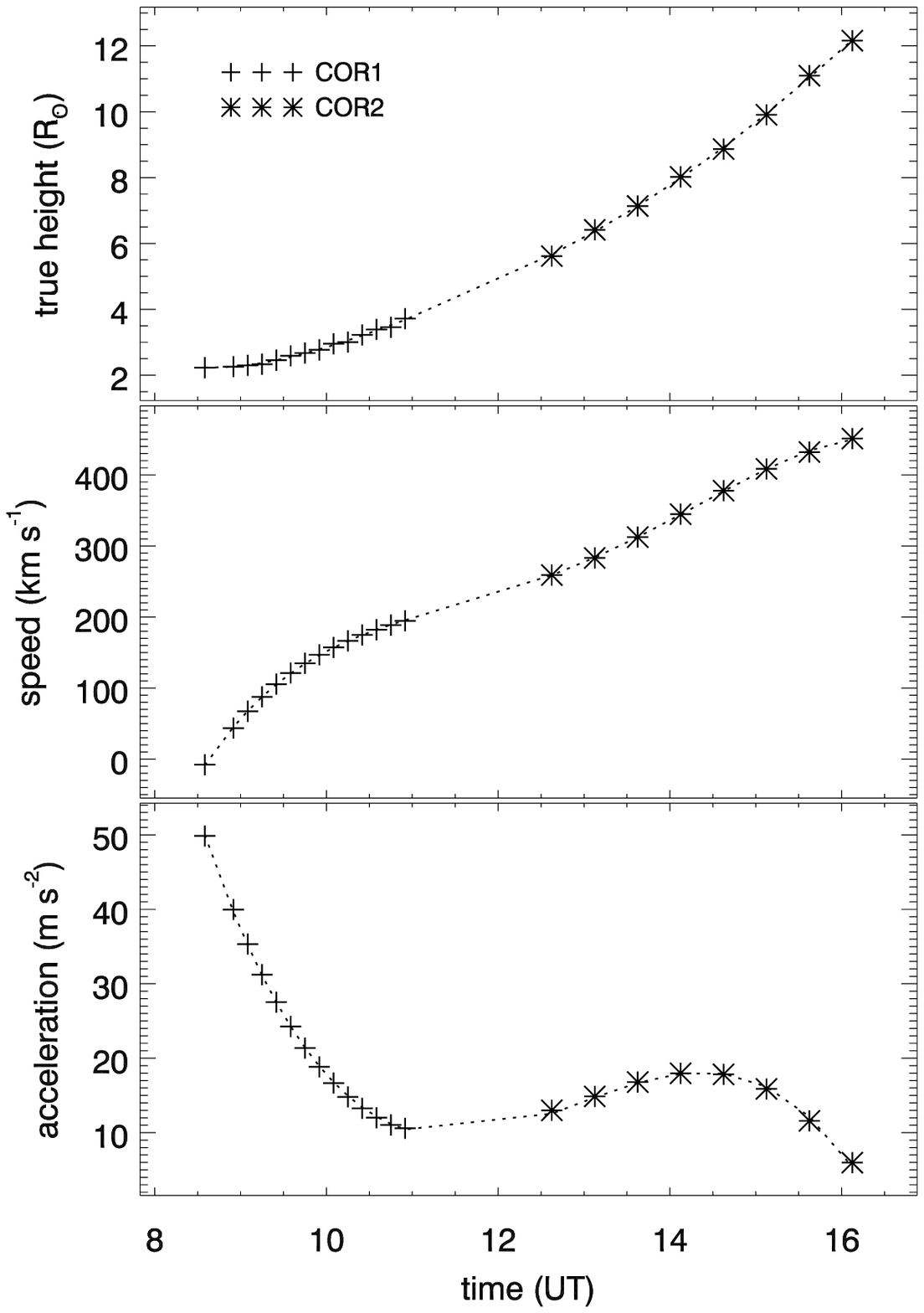}
    \includegraphics[width=0.470\textwidth,clip=]{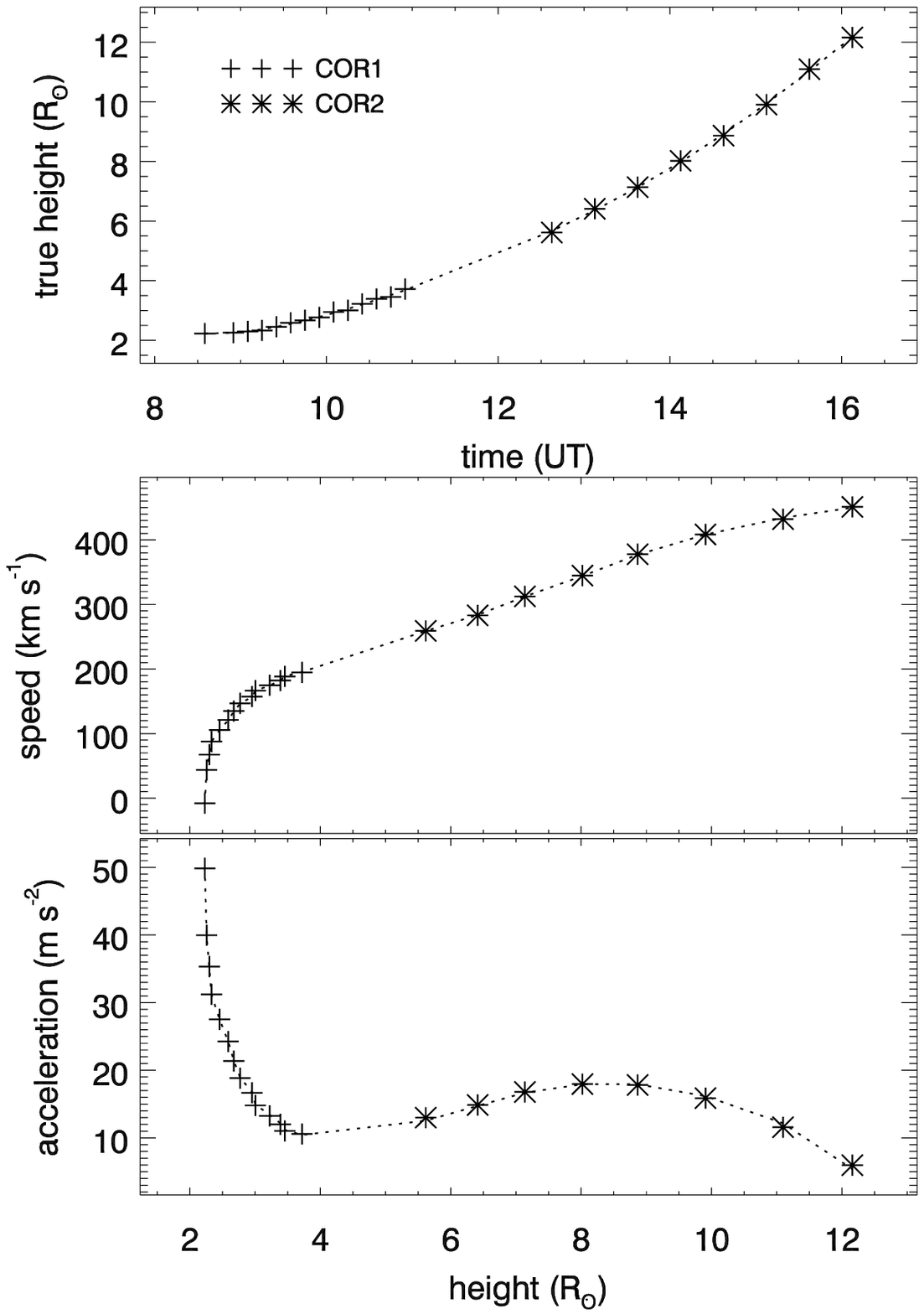}
		}
\caption{Results from the stereoscopic reconstruction applied to a feature in LE of the CME on 2007 November 16. Left: True height of the CME feature against time in the top panel, followed by the true speed and acceleration against time in the middle and bottom panels. Right: True height of the CME against time at the top, followed by the true speed and acceleration against true height in the middle and bottom panels. Plus signs (+) and asterisks (*) indicate that the feature was observed in COR1 and COR2 FOVs, respectively.}\label{F:res16nov}
\end{figure}

\begin{figure}
\centerline
		{
	\includegraphics[width=0.46\textwidth,clip=]{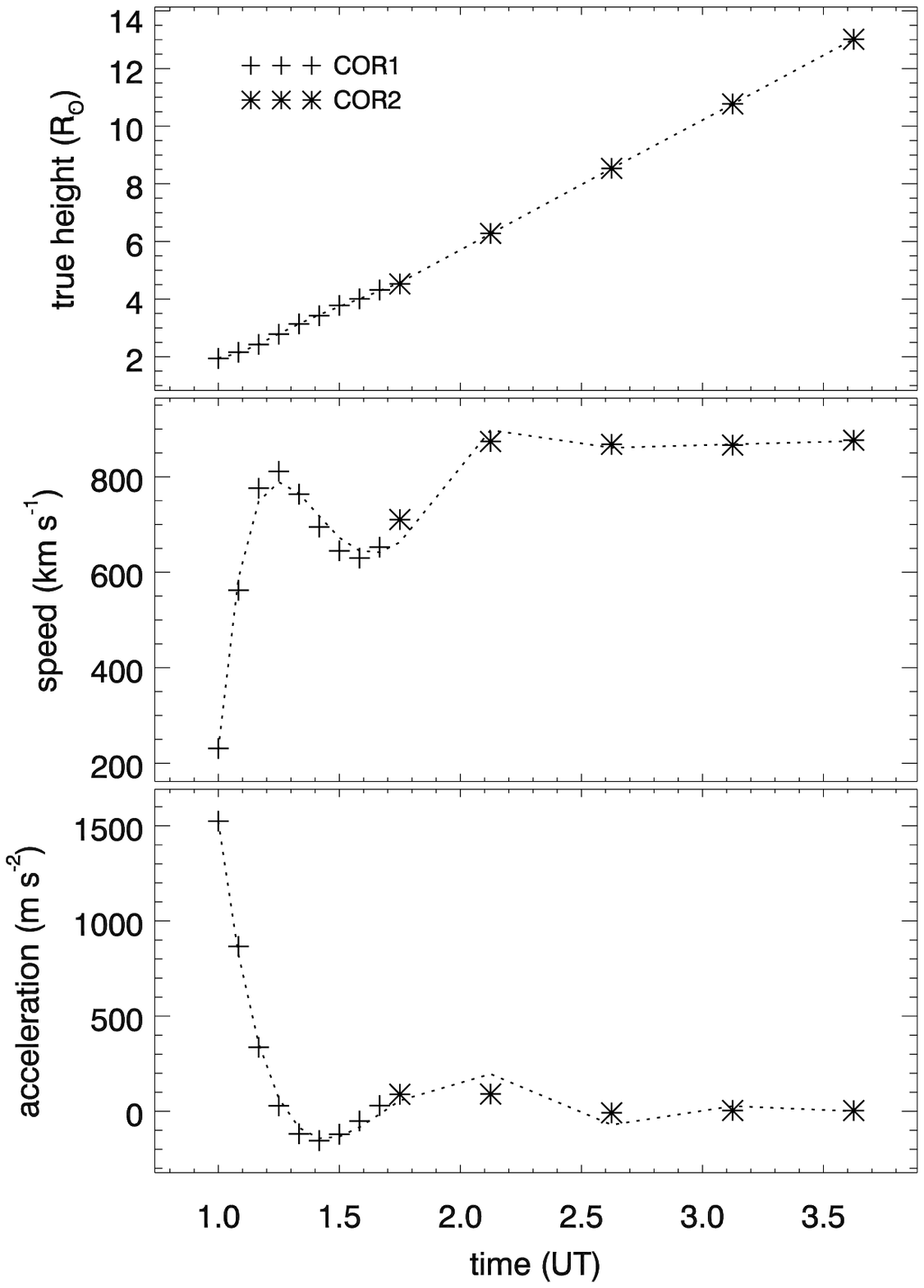}
    \includegraphics[width=0.46\textwidth,clip=]{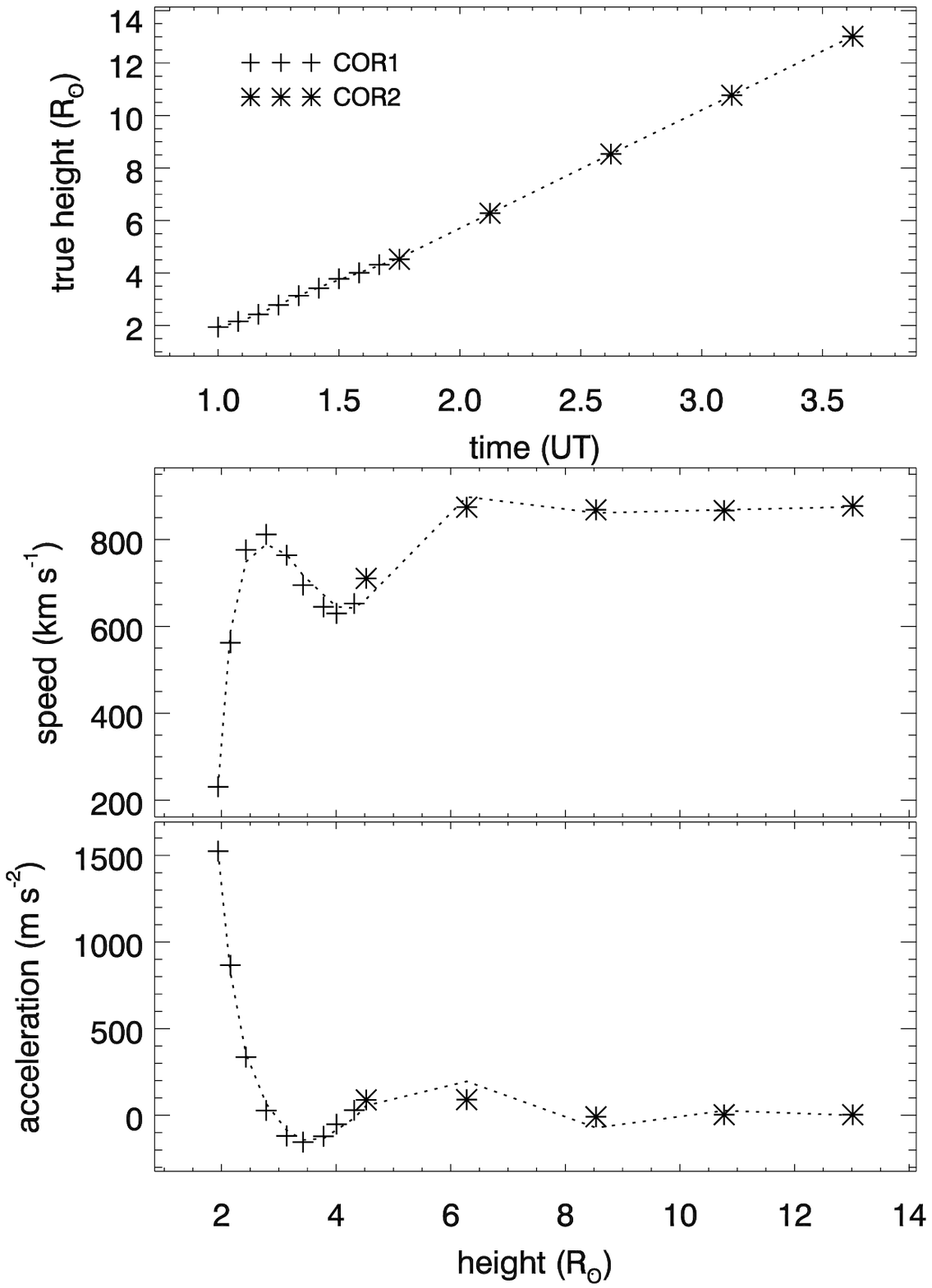}
		}
\caption{Results from stereoscopic reconstruction for CME on 2007 December 31, similar to Figure~\ref{F:res16nov}}\label{F:res31dec}
\end{figure}

\begin{figure}
\centerline
		{
	\includegraphics[width=0.465\textwidth,clip=]{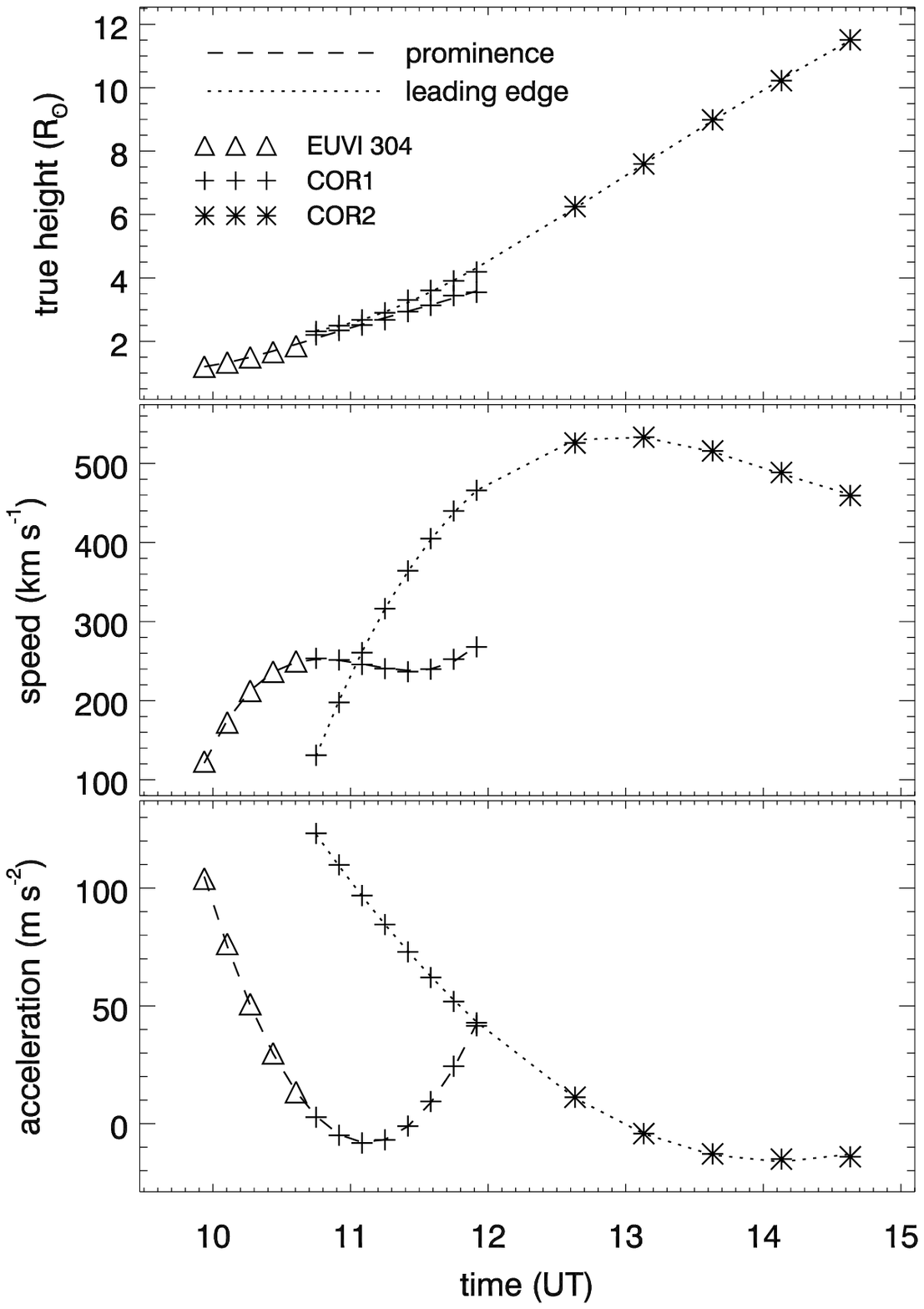}
    \includegraphics[width=0.470\textwidth,clip=]{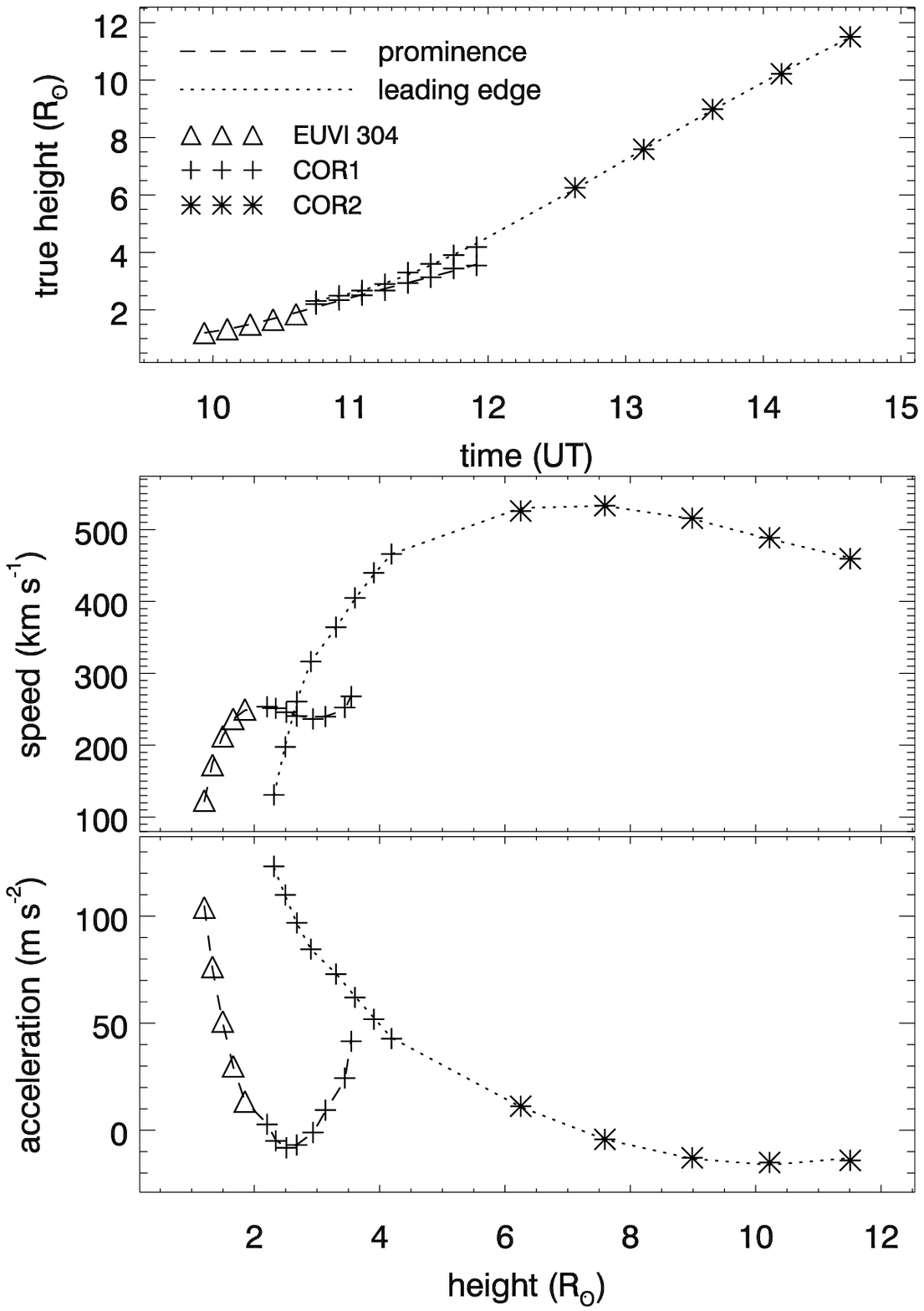}
		}
\caption{Results from stereoscopic reconstruction for CME on 2008 April 9, similar to Figure~\ref{F:res16nov}. Here, triangles ($\vartriangle$) indicate feature observed in EUVI FOV.}\label{F:res09apr}
\end{figure}

\begin{figure}
\centerline
		{
	\includegraphics[width=0.465\textwidth,clip=]{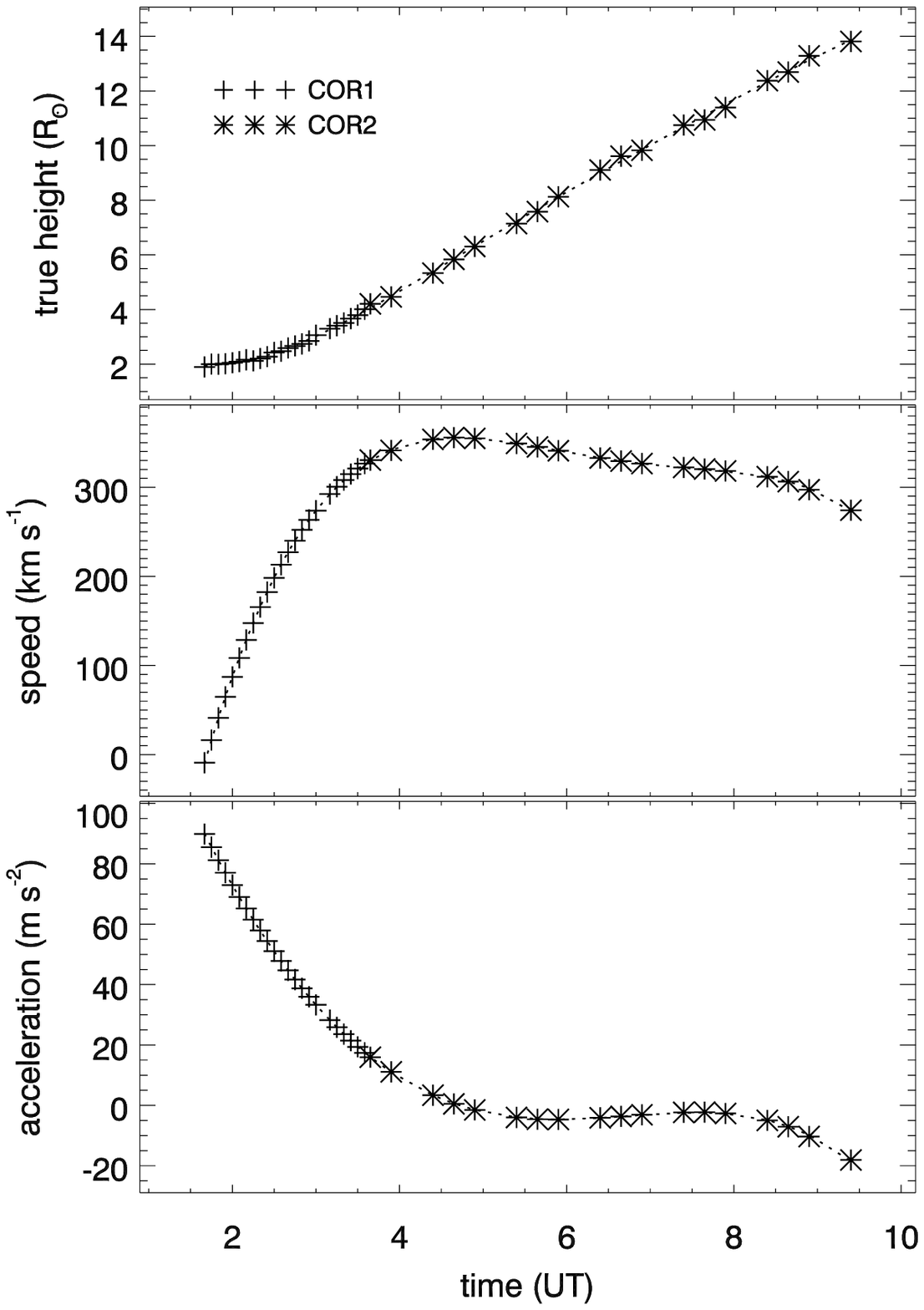}
    \includegraphics[width=0.470\textwidth,clip=]{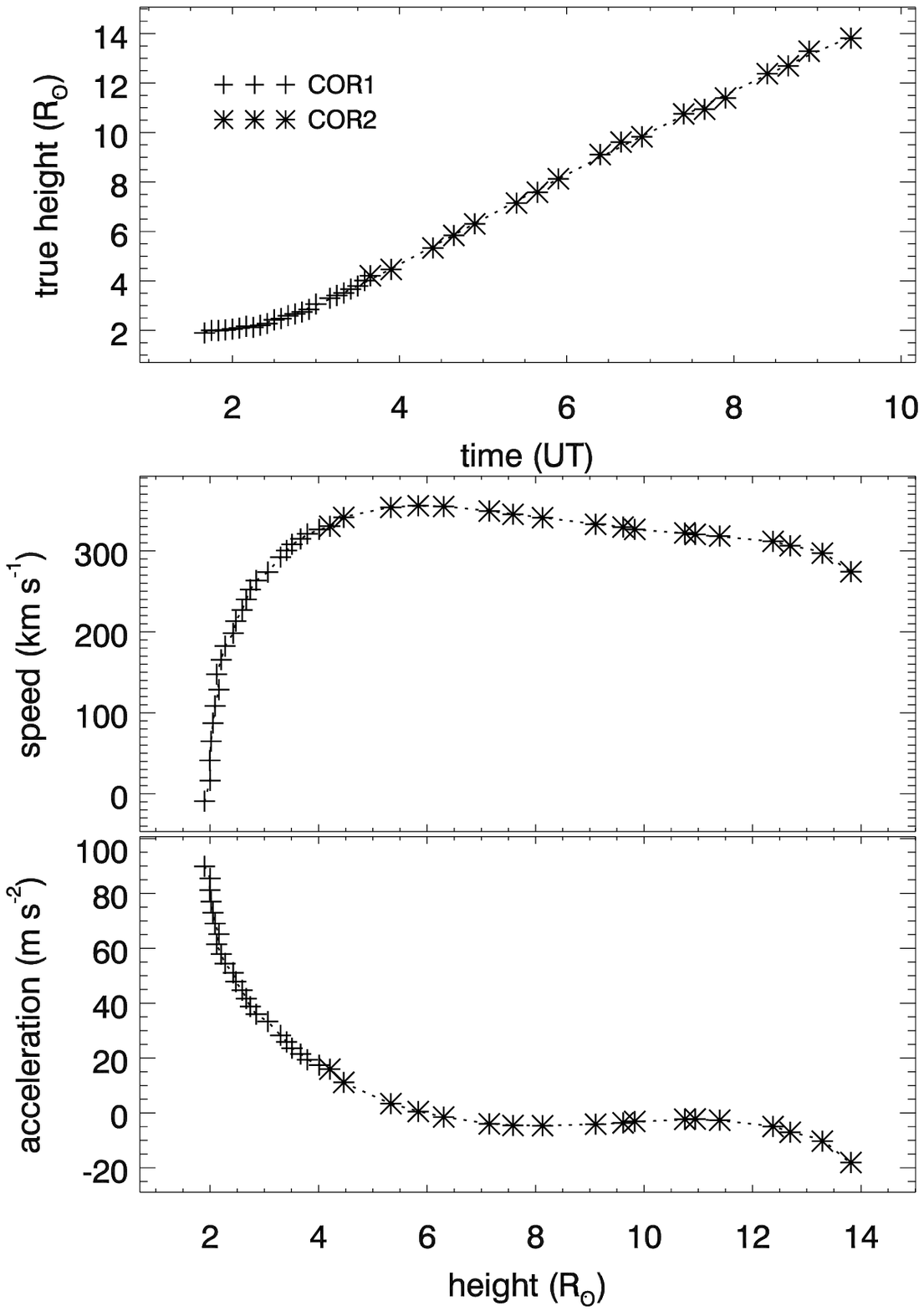}
		}
\caption{Results from stereoscopic reconstruction for CME on 2009 December 16, similar to Figure~\ref{F:res16nov}}\label{F:res16dec}
\end{figure}

\begin{figure}
\centerline
		{
	\includegraphics[width=0.465\textwidth,clip=]{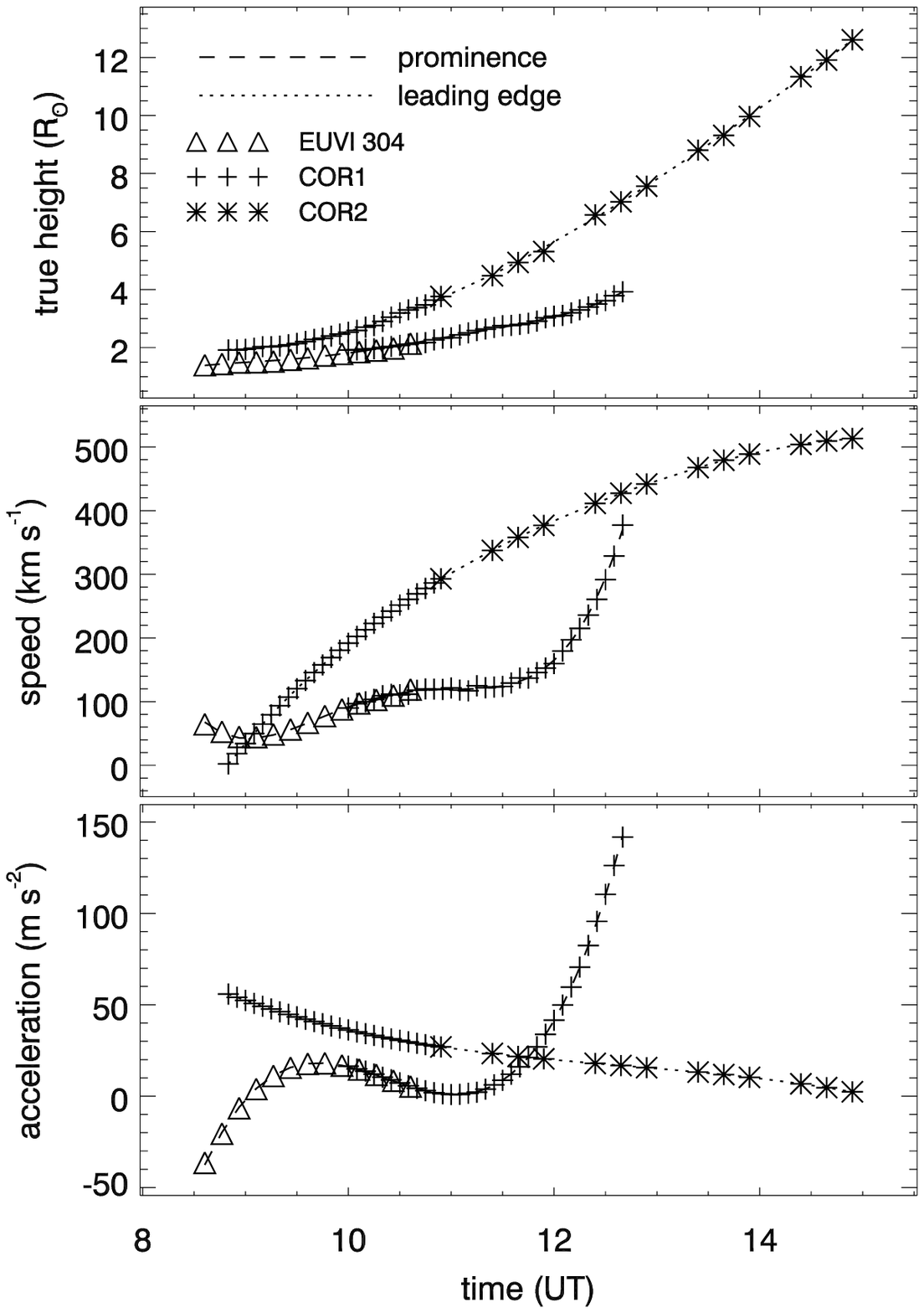}
    \includegraphics[width=0.470\textwidth,clip=]{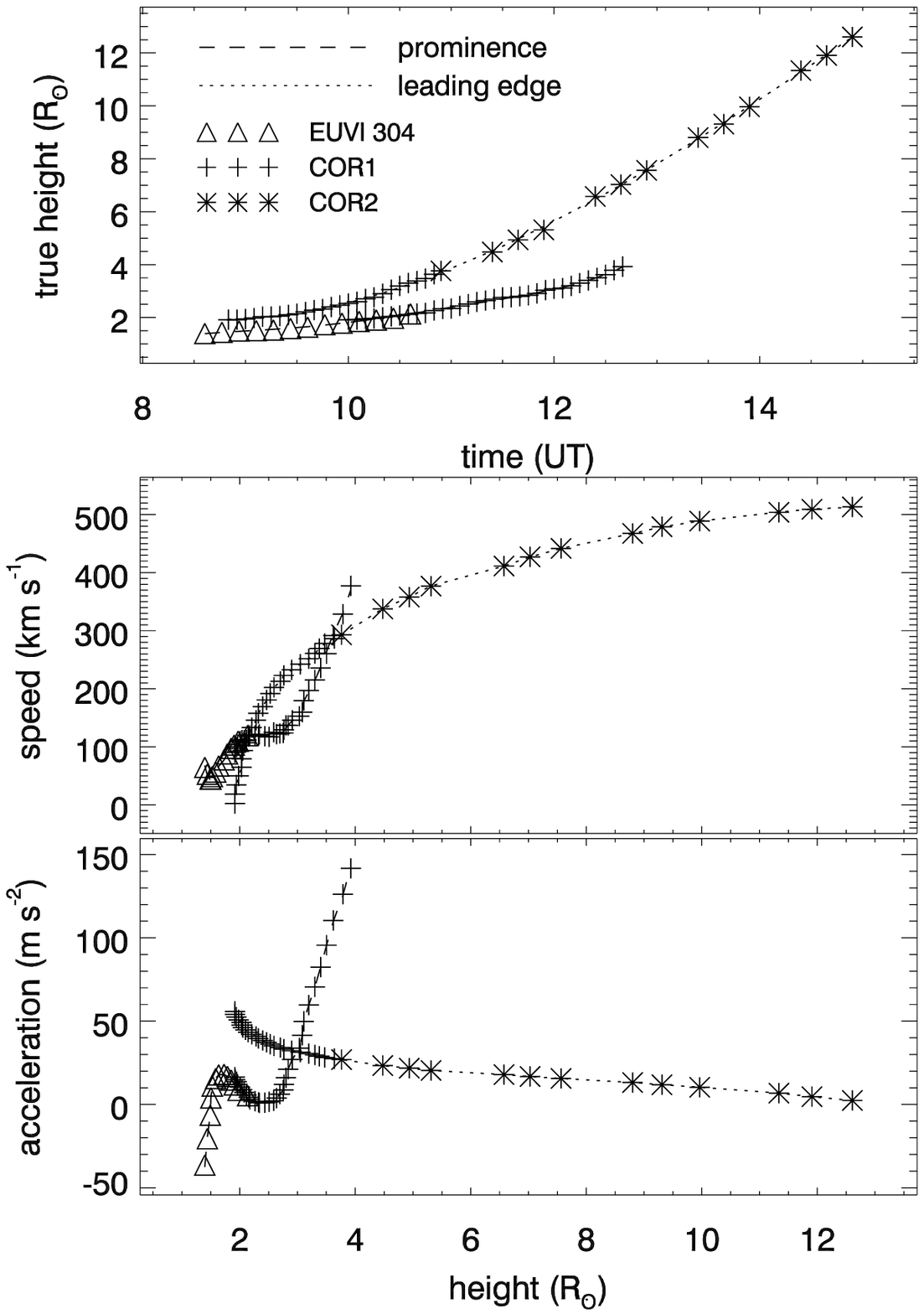}
		}
\caption{Results from stereoscopic reconstruction for CME on 2010 April 13, similar to Figure~\ref{F:res09apr}}\label{F:res13apr}
\end{figure}

\begin{figure}
\centerline
		{
	\includegraphics[width=0.465\textwidth,clip=]{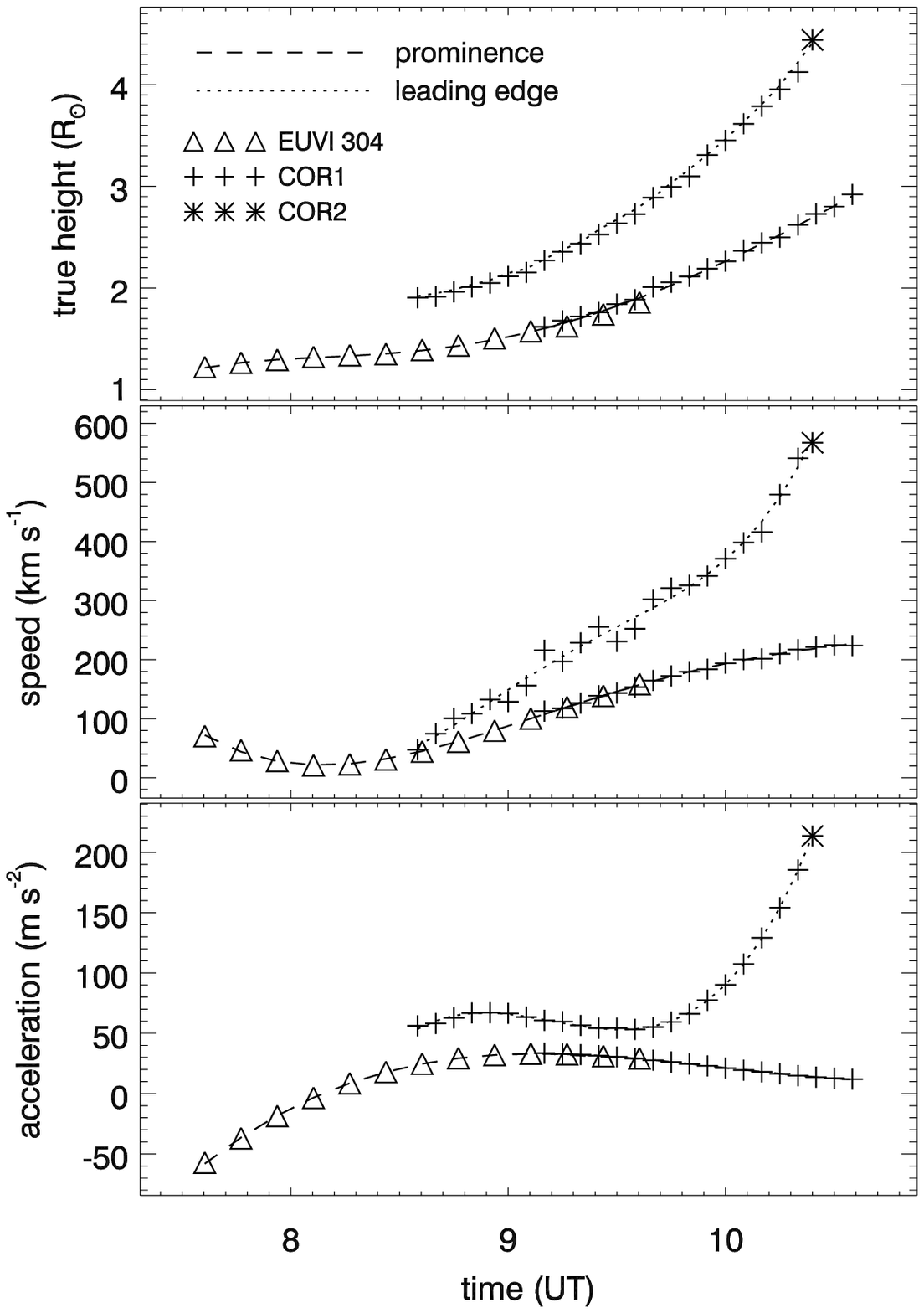}
    \includegraphics[width=0.470\textwidth,clip=]{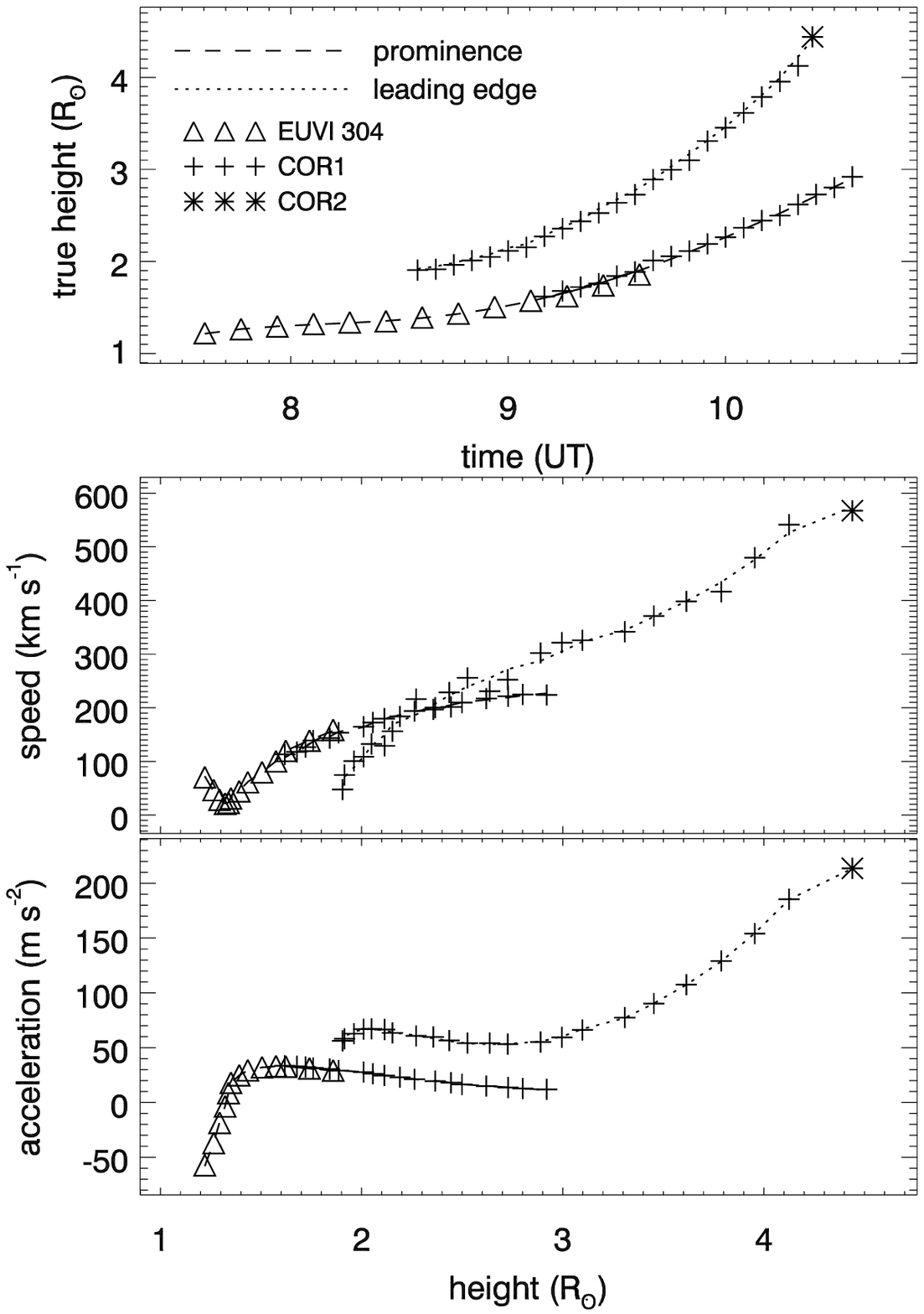}
		}
\caption{Results from stereoscopic reconstruction for CME on 2010 August 1, similar to Figure~\ref{F:res09apr}}\label{F:res01aug}
\end{figure}

We have analysed six CMEs from the coronagraphs COR1 and COR2, and the associated EPs in three of the cases from EUVI on board the identical STEREO\,A and B spacecraft. We identified and tracked a feature in the LE of all the CMEs in both COR1 and COR2, and in the associated prominences, wherever applicable. While most of the earlier studies on CME acceleration were carried out using projected measurements, we have used a stereoscopic reconstruction technique \citep{Joshi.Srivastava2011} to obtain the true coordinates, and hence the true speed and acceleration of the feature. On fitting a polynomial function to the true height, the speed and acceleration of the CMEs as a function of time and true height were determined. The results of the kinematic study of EPs and the CME LEs are shown in Figures~\ref{F:res16nov}--\ref{F:res01aug}. We summarise the results obtained from the reconstruction in Table~\ref{T:summa}.

\vspace{-0.2cm}
\begin{center}
\begin{table}[!tbp]
\begin{tabular}{lrrrrrr}
\hline
Event & $v_{max}$ & height of & $a_{max}$ & height of & $v$ at & $a$ at \\
	  & (\kmps) & $v_{max}$ ($\Rsun$) & (\mpss) & $a_{max}$ ($\Rsun$) & 10$\Rsun$ & 10$\Rsun$ \\
\hline
16 Nov 2007 LE & 451  & 12.2  & 50    & 2.2  & 408  & 16   \\
31 Dec 2007 LE & 876  & 13.0  & 1524  & 1.9  & 860  & 2    \\
~9 Apr 2008 LE & 533  & 7.6   & 123   & 2.3  & 488  & -15  \\
16 Dec 2009 LE & 356  & 5.8   & 90    & 1.9  & 488  & -15  \\
13 Apr 2010 LE & 522  & 12.6  & 61    & 1.9  & 193  & 36   \\
~1 Aug 2010 LE & 567  & 4.4   & 213   & 4.4  & ---  & ---  \\
~9 Apr 2008 EP & 268  & 3.5   & 104   & 1.2  & ---  & ---  \\
13 Apr 2010 EP & 377  & 3.9   & 141   & 3.9  & ---  & ---  \\
~1 Aug 2010 EP & 224  & 2.9   & 34    & 1.6  & ---  & ---  \\
\hline
\end{tabular}
\caption{Summary of the 6 LEs and 3 EPs analysed using three-dimensional reconstruction. $v_{max}$ and $a_{max}$ denote the maximum speed and acceleration of the CME calculated. The heights at which CMEs attained these values are also provided. The last two columns show the speed and acceleration of the CMEs at a distance of $10\Rsun$.}\label{T:summa}
\end{table}
\end{center}
\vspace{-1.0cm}

It is believed that most of the CME acceleration typically occurs in the lower corona. \citet{Chen.Krall2003} have found the height of maximum acceleration of CME to be $2-3\Rsun$ from a study of several CMEs,  \citet{Vrsnak2001a} have considered this height to be $4\Rsun$. However, from our reconstructed results (Figures~\ref{F:res16nov}--\ref{F:res01aug}), we observe that in all the cases studied here, the peak of main phase of acceleration lies below the true height of $2\Rsun$. This indicates that most of the CME dynamics occurs closer to the Sun than previously believed, as shown by \citet{Chen.etal2006} from a comparison of observations and models.

Earlier studies \citep{Zhang.etal2001,Chen.Krall2003}, have observed CMEs in all the three SoHO/LASCO coronagraphs which together cover a range from $1.1-32\Rsun$. In such studies, the initiation phase of the CME, as well as peak of their acceleration could be observed. The COR1 and COR2 coronagraphs together image the solar corona from $1.4$ to $15.0\Rsun$, however these are only the plane-of-sky FOVs of the coronagraphs. The minimum value of true height of reconstructed features the corona is approximately $2.0\Rsun$. Thus, in most of the cases we do not capture the rise phase acceleration of the LE of CME. In all but one case studied here, the acceleration peak has already passed from the time we start observing the CME. At this point, it is necessary to point out that the heights determined in this study are true heliocentric distances, hence they are seen to be significantly different than the heights obtained from previous studies which relied upon observations from a single spacecraft.

The CME on 2007 December 31 was associated with a flare having X-ray class C8, however it still showed a very high value of acceleration of over $1500\mpss$. Earlier studies have shown that the acceleration phase of CMEs coincides with the increase in soft X-ray flux due to the associated flare \citep{Neupert.etal2001,Shanmugaraju.etal2003}. \citet{Maricic.etal2007} have also shown that both the velocity and acceleration of the CME show a significant correlation with the X-ray class of the associated flare. As per the least squares fit obtained from their study, acceleration of the CME associated with a C8 flare should be around $300\mpss$. The value calculated by us however, is 5 times more, suggesting that the flare energy alone might not be the only one to drive the CMEs. In such a scenario, the supposition that impulsive and gradual CMEs are respectively associated with flares and EPs \citep{MacQueen.Fisher1983,Moon.etal2002} should also be subjected to further scrutiny. Also, deviations to the findings reported by \citet{Maricic.etal2007}, where acceleration of CMEs is correlated with the X-ray class of the associated flare, should not be ignored. \citet{Raftery.etal2010} have used soft and hard X-ray observations in addition to STEREO observations \citep{Lin.etal2010} to analyse the 2007 December 31 CME, and have found that it follows the tether-cutting reconnection model.

\begin{figure}
\centerline
		{
	\includegraphics[width=0.48\textwidth,clip=]{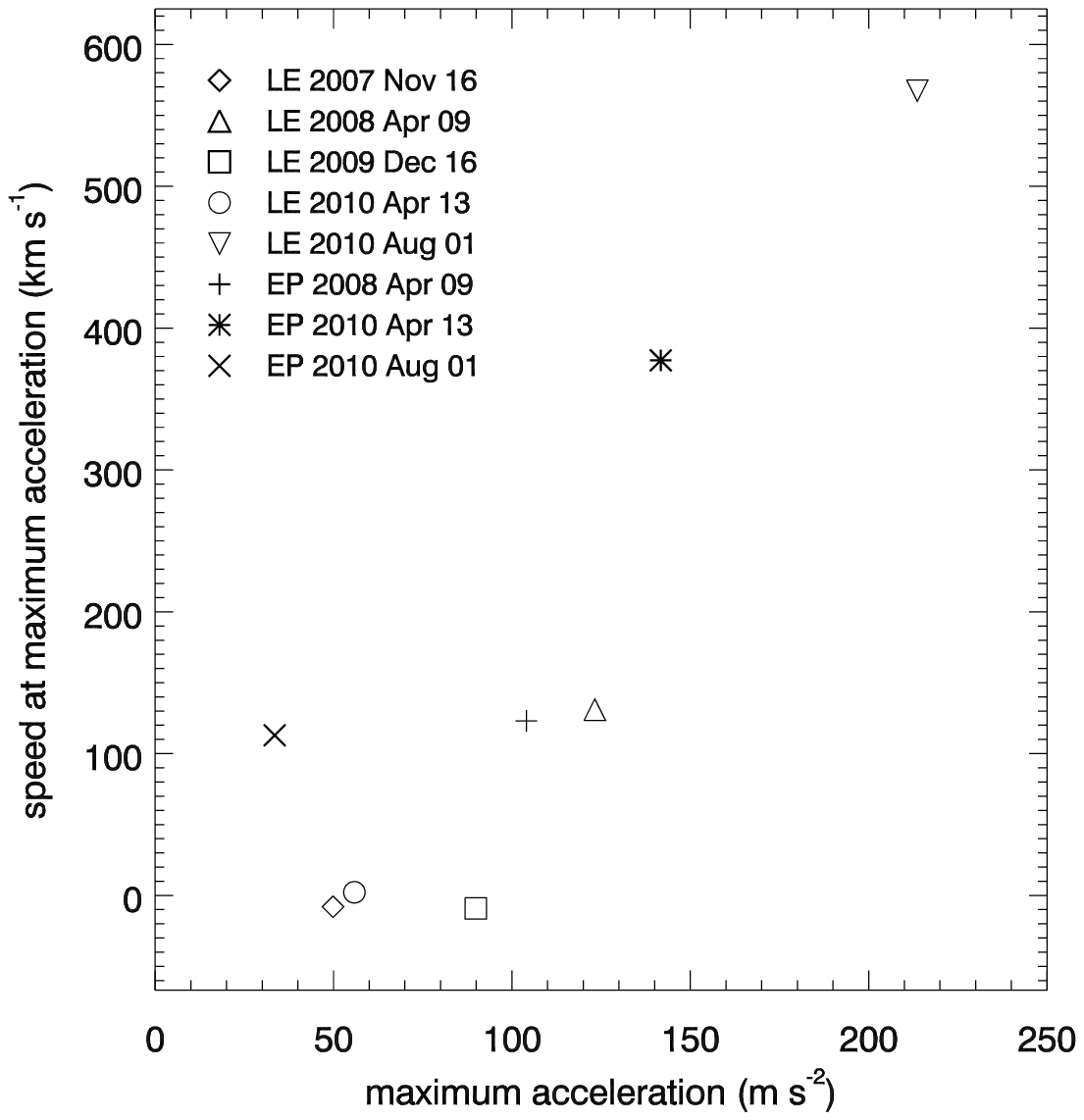}
	\includegraphics[width=0.47\textwidth,clip=]{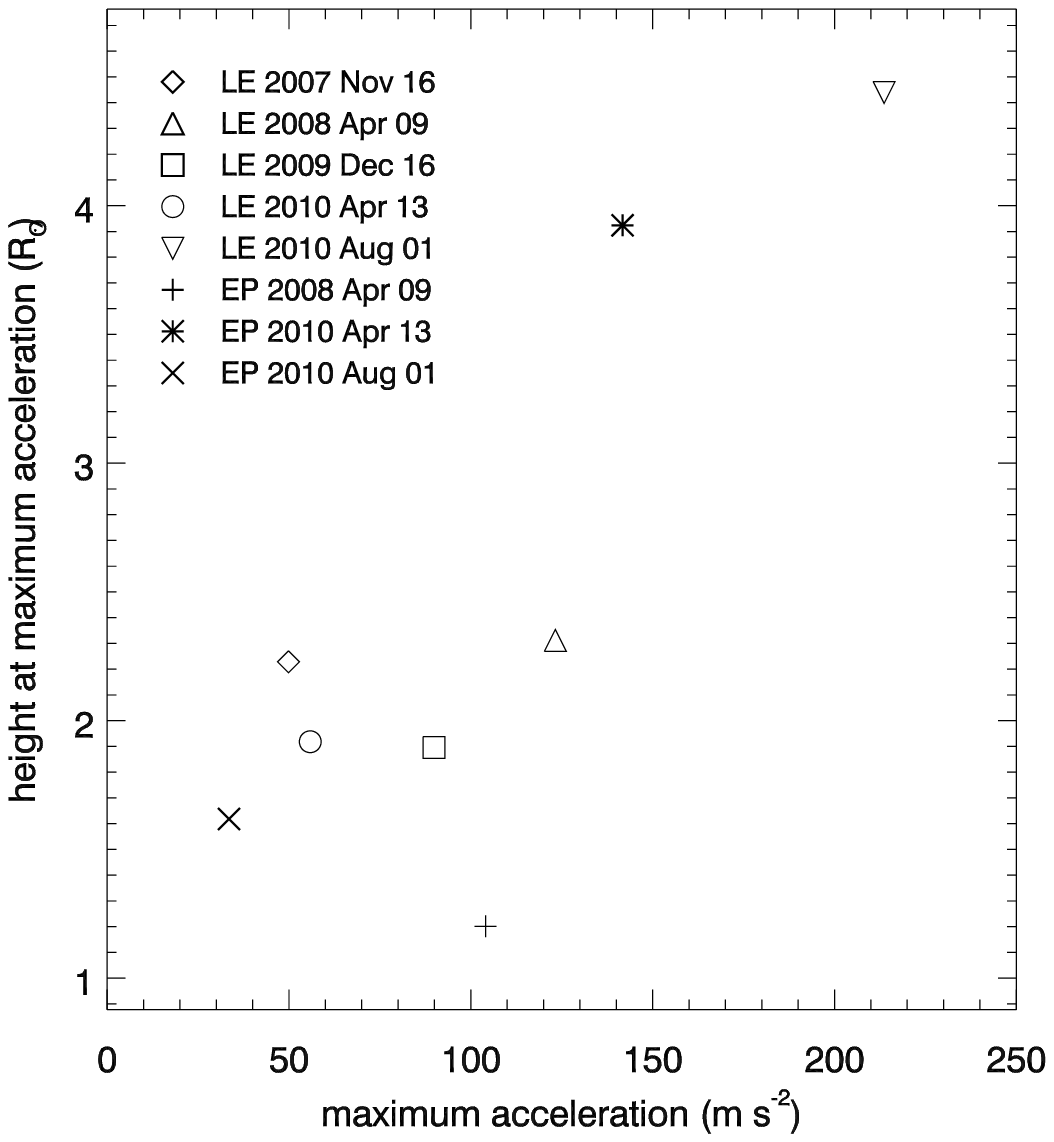}
		}
\caption{Two scatter plot showing speed (left panel), and height (right panel) at the instance of maximum acceleration and the maximum acceleration itself of the 6 CMEs and 3 EPs studied. The legend shows data points corresponding to each event. Data point for 2007 December 31 is not shown since it has a very high value of maximum acceleration.}\label{F:max_spd}
\end{figure}

The results obtained from reconstruction were used to determine maximum acceleration, average acceleration, acceleration magnitude, and acceleration duration, attained by the CMEs and EPs. The time interval between the maximum and the zero value of acceleration is termed as the acceleration duration, while the velocity increase during this time divided by acceleration duration is termed as the acceleration magnitude \citep{Zhang.Dere2006}. The very high value of acceleration for the 2007 December 31 CME makes the event an `outlier', hence, we have not included that value in the scatter plots in Figure~\ref{F:max_spd}. The left panel shows speed at maximum acceleration plotted against the maximum acceleration. From this figure, we find that in the events studied by us, higher the maximum value of acceleration, higher is the speed at that instant. While, the right panel of Figure~\ref{F:max_spd} shows scatter plot of maximum acceleration and height at the instance of maximum acceleration. This scatter plot suggests, that higher the acceleration, higher up in the corona it occurs.

\textbf{Although the acceleration in this study is determined up to the COR2 FOV, it may not be the value with which the CME is travelling at larger distances from the Sun. Based on interplanetary measurements, it was shown earlier by \citet{Gopalswamy.etal2000} that slow CMEs tend to accelerate, while the faster ones tend to decelerate. Recently, \citet{Davis.etal2010} have measured the speeds of 26 CME from the Heliospheric Imagers (HI), which are part of the SECCHI suite on the STEREO spacecraft. In their study, they have found that CMEs with speeds less than $400\kmps$ in COR2 FOV have higher speeds in HI FOV, and vice-versa. Thus they have cautioned that a CME may undergo genuine acceleration even in the HI FOV, which extend from $15-84\Rsun$ for HI-1 and $66-318\Rsun$ for HI-2.}
 
Previous studies have reported that acceleration of a CME shows bimodal distribution \citep{Chen.Krall2003}. We observe such a bimodal distribution in 3 CMEs, the ones which are not associated with prominence eruptions. The residual acceleration for the very impulsive 2007 December 31 CME was $90\mpss$, while for the CMEs on 2007 November 16 and 2009 December 16, it was found to be $18\mpss$ and $-2\mpss$, respectively. The other CMEs, which are associated with prominences do not show such an acceleration profile. \citet{Chen.Krall2003} have invoked the flux injection mechanism to trigger an eruption in a magnetic flux rope, which leads to the residual acceleration phase. In the cases analysed here, we observe that only the flare-associated CMEs undergo residual acceleration, which indicates that flux injection seems to be a good explanation for eruption of the flare-associated CMEs studied here, but a different mechanism should be considered for EP-associated CMEs.

Of the three CMEs associated with prominences, the 2010 April 13 and 2010 August 1 were associated with large quiescent polar crown prominences, while the one on 2008 April 9 was associated with an active-region prominence. We find that the prominences on 2008 April 9 and 2010 April 13 showed a strong positive acceleration in the COR1 FOV, when their heights were close to almost $4\Rsun$. During the same time however, acceleration of the CME LE was decreasing. This indicates that even at a height of $4\Rsun$, forces acting on the CME and the EP cannot be considered to be the same, as suggested by \citet{Srivastava.etal2000} and \citet{Maricic.etal2004}.

Thus, in this study, from the 3D reconstruction of six CMEs and EPs associated with three of them, we have observed some aspects of their acceleration, as detailed above, which were not previously reported. We find that the maximum CME acceleration occurs at a height of less than $2\Rsun$, where earlier, this height was believed to be between $2-4\Rsun$. The bimodal acceleration profile was not observed in EP-associated CMEs, but in only those CMEs that were not associated with EPs. Two of the three prominences in the study showed a high and rising value of acceleration at a distance of almost $4\Rsun$ but the corresponding CME LE does not show the same behaviour. The CME on 2007 December 31, showed acceleration of over $1500\mpss$, which is unusually high for a CME associated with a C-class flare.

The authors thank the STEREO/SECCHI consortium for providing the data. The SECCHI data used here were produced by an international consortium of the Naval Research Laboratory (USA), Lockheed Martin Solar and Astrophysics Lab (USA), NASA Goddard Space Flight Center (USA), Rutherford Appleton Laboratory (UK), University of Birmingham (UK), Max-Planck-Institut for Solar System Research (Germany), Centre Spatiale de Li$\grave{\textrm e}$ge (Belgium), Institut d'Optique Theorique et Appliqu$\acute{\textrm e}$e (France), Institut d'Astrophysique Spatiale (France). Work by N.S. partially contributes to the research on collaborative NSF grant \mbox{ATM-0837915} to Helio Research.

\end{document}